\documentclass[prd,tightenlines,nofootinbib,superscriptaddress,twocolumn]{revtex4}

\usepackage{amsfonts,amssymb,amsthm,bbm,amsmath}
\usepackage{hyperref}
\usepackage{color,psfrag}
\usepackage[dvips]{graphicx}
\usepackage{tabularx}
\usepackage{multirow}	
\usepackage{eurosym}
\usepackage{color}
\usepackage[dvipsnames]{xcolor}
\usepackage{fancyhdr}
\usepackage{fancybox}
\usepackage{stmaryrd} 
\usepackage{slashed}
\usepackage{dsfont}
\usepackage{lastpage}
\usepackage{refcount}
\usepackage{CJKutf8}

\usepackage{todonotes}

\usepackage{tikz}
\usepackage{circuitikz}
\usetikzlibrary{shapes.geometric,calc,intersections,
decorations.markings,decorations.text,decorations,patterns, arrows,
decorations.pathmorphing,fpu,math,babel}

\usepackage{subcaption}
\newcommand{\me}{\mathrm{e}}
\newcommand{\mi}{\mathrm{i}}
\newcommand{\md}{\mathrm{d}}

\DeclareMathOperator{\trace}{tr}

\newcommand{\ket}[1]{{\left|#1\right\rangle}}
\newcommand{\bra}[1]{{\left\langle#1\right|}}
\newcommand{\dket}[1]{{\left|#1\right]}}
\newcommand{\dbra}[1]{{\left[ #1\right|}}

\newcommand{\C}{{\mathbb C}}
\newcommand{\N}{{\mathbb N}}
\newcommand{\R}{{\mathbb R}}

\newcommand{\cH}{{\mathcal H}}

\newcommand{\cR}{{\mathcal R}}

\newcommand{\cP}{{\mathcal P}}

\newcommand{\cS}{{\mathcal S}}

\newcommand{\cX}{{\mathcal X}}
\def\cY{{\cal Y}}

\newcommand{\SU}{\mathrm{SU}}

\newcommand{\SL}{\mathrm{SL}}
\newcommand{\SO}{\mathrm{SO}}
\newcommand{\Sp}{\mathrm{Sp}}
\newcommand{\U}{\mathrm{U}}

\newcommand{\be}{\begin{equation}}
\newcommand{\ee}{\end{equation}}
\newcommand{\beq}{\begin{eqnarray}}
\newcommand{\eeq}{\end{eqnarray}}
\newcommand{\bes}{\begin{eqnarray}}
\newcommand{\ees}{\end{eqnarray}}

\newcommand{\mat} [2] {\left ( \begin{array}{#1}#2\end{array} \right ) }

\newcommand{\su}{{\mathfrak su}}

\renewcommand{\u}{{\mathfrak u}}

\newcommand{\so}{{\mathfrak so}}

\newcommand{\la}{\langle}
\newcommand{\ra}{\rangle}

\newcommand{\tr}{{\mathrm{Tr}}}
\newcommand{\f}{\frac}

\def\nn{\nonumber}
\def\pp{\partial}

\def\eps{\epsilon}

\newcommand{\id}{\mathbb{I}}

\def\om{\omega}
\def\tX{\widetilde{X}}
\def\tcX{\widetilde{\cX}}
\def\vC{\vec{C}}
\def\vJ{\vec{J}}

\def\vR{\vec{R}}

\def\vsigma{\vec{\sigma}}

\def\bF{\bar{F}}

\def\vX{\vec{X}}
\def\vY{\vec{Y}}
\def\vZ{\vec{Z}}
\def\bgamma{\bar{\gamma}}
\def\hX{\hat{X}}
\def\bz{\overline{z}}
\def\he{\hat{e}}

\begin{document}

\title{Quantum Surface and Intertwiner Dynamics in Loop Quantum Gravity}

\author{{\bf Alexandre Feller}}\email{alexandre.feller@ens-lyon.fr}
\affiliation{Univ Lyon, Ens de Lyon, Université Claude Bernard Lyon 1, CNRS, 
Laboratoire de Physique, F-69342 Lyon, France}

\author{{\bf Etera R. Livine}}\email{etera.livine@ens-lyon.fr}
\affiliation{Univ Lyon, Ens de Lyon, Université Claude Bernard Lyon 1, CNRS, 
Laboratoire de Physique, F-69342 Lyon, France}

\date{\today}

\begin{abstract}


We introduce simple generic models of surface dynamics in loop quantum gravity (LQG). A quantum surface is defined as a set of elementary patches of area glued together.
We provide it with an extra structure of locality (nearest neighbors), thought of as induced by the whole spin network state
defining the 3d bulk geometry around the quantum surface.
Here, we focus on classical surface dynamics, using a spinorial  description of surface degrees of freedom.
We introduce two classes of dynamics,  to be thought as templates for future investigation of LQG dynamics with in mind the dynamics 
of quantum black holes.
The first defines global dynamics of the closure defect of the surface, with two basic toy-models, either a dissipative dynamics relaxing towards the closure constraint or a Hamiltonian dynamics precessing the closure defect.
The second class of dynamics describes the isolated regime, when both area and closure defect are conserved throughout the evolution. The surface dynamics is implemented through $\U(N)$ transformations and generalizes to a Bose-Hubbard Hamiltonian with a local quadratic potential interaction.
We briefly discuss the implications of modeling the quantum black hole dynamics by a surface Bose-Hubbard model.

\end{abstract}

\maketitle

\section*{Introduction}

A guiding idea for quantum gravity is the holographic principle. It claims that the dynamics of the bulk geometry can be projected onto its boundary and that the whole theory can be described through degrees of freedom living on such holographic screens.
This principle deeply intertwines with the coarse-graining of geometry and renormalization of its dynamics in quantum gravity. For instance, considering a small bounded region of the 3d space which is to coarse-grained to a single point (in a renormalisation group scheme \`a la Wilson), the dynamics of the bulk geometry (within that region) can be projected onto its boundary and described through degrees of freedom living on that boundary surface. These surface degrees of freedom contain all the relevant data for the interaction of that 3d region with the exterior. Then the dynamics of that surface will be understood as the renormalized dynamics of local effective degrees of freedom living at the coarse-grained point.

This underlines the importance of studying the surface dynamics in quantum gravity. At the classical level, this leads to a reformulation of general relativity in hydrodynamic and thermodynamic terms. Indeed, when considering a $2d$ surface (null, space-like or time-like), the Einstein equations reduce to the  $2d$ Navier-Stokes equation and physical quantities such as viscosity or surface charge density 
can be defined and encode various components of the curvature tensor. This point of view has been particularly developed for black hole  leading to the membrane paradigm for black hole horizons \cite{Damour_1982,Thorne_1986,Freidel_2015}.  Reciprocally, it is possible to reverse this logic and reconstruct all of general relativity from the (thermo)dynamics of surfaces and boundaries \cite{Jacobson:1995ab,Padmanabhan:2013nxa,Padmanabhan:2015zmr}.
Following this logic, understanding surface dynamics  in the quantum regime appears to be a crucial task.

Loop quantum gravity (LQG) is a proposal for a full non perturbative quantum theory 
of gravity initially based on a canonical quantization of general relativity  and later developed as a path integral in the spinfoam framework.
(for textbooks, see \cite{Rovelli_book, Rovelli_Vidotto_book,Thiemann_book}).
A discrete geometry picture emerges from the analysis of geometrical operators
and the quantum geometry is described by spin network states.
Reconstructing a classical geometry, with a metric and local coordinates, from those quantum states becomes rather subtle due to 
background independence and diffeomorphism invariance.
In this context, a quantum surface is defined as set of elementary patches 
of quanta of geometry glued together.
Moreover, we introduce here a notion  of surface locality, describing how those patches are glued together to form the surface.
In a classical setting, a coordinate system and metric in the bulk 3d will induce a coordinate system and metric on the boundary surface. Similarly, at the quantum level, locality on the surface should be induced by the whole structure of the spin network, which defines an embedding of the surface in 3d space. For instance, we would think of locality on a black hole horizon induced by the near-horizon geometry.
In the present work, we will forget the bulk structure, retain only the locality that it induces on the quantum surface and focus on defining and analyzing the surface dynamics  per se. 

Ideally, the dynamics of a quantum surface would be given by solving the Hamiltonian 
constraint of the theory. Since the implementation of the constraint algebra as quantum operators is still under 
active research, we adopt a more indirect point of view. We  focus on a 
classical analysis, describing the surface classically with a spinor phase 
space representation \cite{Livine_2013,Livine:2013gna}  
and define generic models of surface  dynamics, in terms of Hamiltonians polynomial in the spinor variables, that would serve 
as templates for quantum dynamics in LQG.
Since the spinor variables directly labels coherent spin network states, these models of classical dynamics can be further understood as defining the dynamics of coherent states of the  quantum geometry 
\cite{Freidel:2010aq,Freidel:2010bw,Borja:2010rc,Freidel_Livine_2011,Livine:2011gp}.

We will consider different types of surface dynamics. We place ourselves in the logic of the renormalization of quantum gravity, where a fundamental dynamics leads to effective dynamics with various new terms, emerging at different scales and  exploring various modes of deformation of the (boundary) surface geometry. We investigate two classes of dynamics. On the one hand, we consider global surface dynamics. In this case, we scrub the notion of locality on the surface and consider all the elementary surface patches on the same footing. This is in line with the point of view of coarse-graining spin networks: a bounded 3d region is coarse-grained to a single spin network vertex, dressed with a closure defect \cite{Livine:2013gna,Charles_2016} which accounts for the curvature excitations within the region. Our model of global surface dynamics will focus on the dynamics of the closure defect associated to the surface. We will introduce two different templates: a dissipative model\footnotemark where the closure defect relaxes towards the closure constraint (vanishing closure defect) and a Hamiltonian model of precession of the closure defect.
\footnotetext{In the context of a surface, which is not isolated, it is natural to introduce dissipation due to its interaction with the bulk geometry, interpreted as the environment. At the quantum level, this leads to considering potential decoherence phenomena of the surface degrees of freedom, when tracing out the fluctuations of the bulk geometry (inside or outside the surface) as discussed in \cite{Feller:2016zuk}.}
The second class of dynamics  deals with local surface dynamics. The notion of locality materializes as the organisation of the surface patches into a network with nearest-neighbor interactions. We will introduce a generic model of isolated dynamics, where both the surface area and closure defect are invariant, defined by a Bose-Hubbard Hamiltonian with a local potential and a hopping term leading to area fluctuations on the surface. This model has a rich phase structure in the quantum regime, which will be interesting to apply to black hole horizons in loop quantum gravity.

A later step will be to work out the quantum dynamics. Spinors are quantized to $\SU(2)$ irreducible 
representations -spins- and  the spinor phase space becomes tensor product of spins forming discrete quantum surfaces.
Then the dynamics we introduce here could then be pushed to the quantum regime, which we leave for future investigation.

The paper is structured as follows. Section \ref{PhaseSpace} reviews
the basic mathematical tools, namely the spinor phase space for surfaces, to construct the dynamics. 
Section \ref{GlobalDyn} presents models of global dynamics of the quantum surface, focusing on the evolution of the area and of the closure defect, with special emphasis on a model of dissipative dynamics relaxing to closure constraint. Section \ref{LocalDyn} investigate the definition of local dynamics on the quantum surface, focusing on an isolated regime when surface deformations  leave both the total area and the closure defect invariant during the evolution. In particular, we will introduce a classical Bose-Hubbard model as a generic template, with possible application to quantum black hole horizons.
Finally  Section \ref{Conclusion} concludes with a discussion of the phase diagram for quantum surfaces in loop quantum gravity induced by the various dynamics we present.

\section{Surface Phase Space in Loop Gravity}
\label{PhaseSpace}

In loop quantum gravity, the quantum 3D geometry is defined as a spin network state. This is a graph\footnotemark{} dressed with algebraic data: $\SU(2)$-representations - spins - on the edges and intertwiners on the vertices. It can be interpreted as a discrete geometry, more precisely as a twisted geometry \cite{Freidel:2010aq,Freidel:2010bw,Dupuis:2012yw,Freidel:2013bfa}, with the spins defining quanta of area and intertwiners defining elementary chunks of volume.
\footnotetext{Rigorously, a spin network state is defined as a projective limit of graphs\cite{Ashtekar:1994mh}, that is one considers a graph as embedded in all possible finer graphs containing it as a subgraph. This allows to properly define  a superposition of spin network states living on a priori different graphs, but thought as states living on a finer graph containing all the graphs of the considered superposition.}
The graph is a priori non-embedded and the spin network defines itself the 3d geometry, which must be constructed from both the combinatorial data of the graph and the algebraic data living on it\footnotemark.
\footnotetext{There exists subtleties on the mathematical definition of spin networks. One can very well strictly work with  non-embedded spin networks, but one can introduce further refinements of topological nature. For instance, one can consider equivalence classes under 3d diffeomorphisms of graphs embedded in a 3d manifold, as in the original definition of loop quantum gravity, or spin network enhanced with topological data \cite{Denicola:2010ni,Duston:2011gk}.
}

Each edge $e$ of the spin network defines an elementary surface, thought as transversal to the edge, and the half-integer spin carried by the edge $j_{e}\in\N/2$ gives the quantum of area carried by that surface in Planck units. The fundamental structures of spin network states being those elementary surface states, it appears to be a very natural framework to provide a proper description of quantum surfaces at both kinematical and dynamical levels and thus to implement the holographic principle in quantum gravity.

\subsection{Quantum Surfaces as Boundaries}


Let us consider a spin network based on a graph $\Gamma$. We consider a bounded region of space, that is a finite set of vertices together with all the edges linking them to each other. We can define such a region by considering an arbitrary embedding to the graph into the Euclidean 3d space and choosing the vertices within a region of the 3d space with the topology of a 3-ball. Then the boundary surface of the region is defined as the set of edges linking one of the region's vertex with an exterior vertex.

For a surface punctured by $N$ edges, the quantum surface Hilbert space is the tensor product of $N$ spins:
\be
\cH_{\cS}=\bigoplus_{\{j_{e}\}_{e=1..N}}\bigotimes_{e=1}^{N}V_{j_{e}}\,
\ee
where $V_{j}$ is the Hilbert space of dimension $\dim V_{j}=d_{j}=(2j+1)$ associated to the spin-$j$ representation. The quantum surface is made of $N$ elementary surface patches, each corresponding to a single edge and carrying a spin $j_{e}$. This spin gives the elementary surface area\footnotemark{} as $a_{e}=\gamma j_{e} l_{P}^{2}$ in Planck units with Immirzi parameter $\gamma$, and the quantum state of that elementary surface patch lives in $V_{j_{e}}$. Above, we have summed all possible spins attached the $N$ edges making the boundary surface.
\footnotetext{Another area spectrum for loop quantum gravity is given by the square-root of the $\SU(2)$ Casimir operator, $a_{e}=\gamma \sqrt{j_{e}(j_{e}+1)} l_{P}^{2}$ and differs from the simpler prescription we took by an operator ordering.}

We have a natural action of $\SU(2)$ on that Hilbert space $\cH_{\cS}$ with group elements acting simultaneously on all the surface patches $V_{j_{e}}$. From the point of view of the spin network, this is a little bit more subtle to come by. Each vertex of the spin network carries an action of $\SU(2)$, which corresponds to local change of frame (thought as gauge transformations). The various surface patches making the whole boundary are a priori not linked to the same vertex and thus are ``desynchronized''. To synchronize their frames and have a single $\SU(2)$ action on the whole boundary surface, one needs to gauge-fix the bulk and effectively reduce it to a single vertex. Following the procedure described in \cite{Livine:2013gna,Charles_2016}, one chooses a root vertex within the region and a maximal tree of the subgraph within the region: that tree will define unique paths from the root vertex to every boundary edge. Gauge-fixing the $\SU(2)$ holonomies to the identity on the tree edges will collapse the bounded region to a single vertex and the $\SU(2)$ action  around the root vertex will be directly transported to the boundary surface: this defines a common reference frame for all the surface patches on the whole surface. In some sense, the tree defines the rigid bulk structure supporting the boundary surface and identifies a growth process from the root vertex in the bulk and the dual surface surrounding it to the boundary surface of the whole region, thereby ensuring that the final surface is closed.
%

Decomposing the surface Hilbert space  $\cH_{\cS}$ into irreducible representations of that global $\SU(2)$ action, we write:
\be
\cH_{\cS}
=
\bigoplus_{J\in\N}
\cR^{(N)}_{J}
=
\bigoplus_{J\in\N}\bigoplus_{C\in\N/2}
\cR^{(N)}_{J,C}\,,
\ee
\be
\textrm{with}\quad
\cR^{(N)}_{J}
=
\bigoplus_{\sum_{e}j_{e}=J}\bigotimes_{e=1}^{N}V_{j_{e}}\,,\quad
\cR^{(N)}_{J,C}
=
\cP_{C}\big{[}
\cR^{(N)}_{J}
\big{]}
\,.
\nn
\ee
$J$ is the sum of all the spins $j_{e}$ and gives the total surface area. $C$ is called the closure defect. It is the spin to which all the spins $j_{e}$ recouple. It is the spin of the global $\SU(2)$ action on the surface, meaning that if we call $\vJ^{(e)}$ the $\su(2)$ generators acting on each elementary surface patch, then their sum $\vJ$ is of norm $C$:
\be
\vJ=\sum_{e}\vJ^{(e)}\,,
\quad
\vJ^{(e)}{}^{2}=j_{e}(j_{e}+1)\,,
\quad
\vJ^{2}=C(C+1)\,.
\ee
The projector  $\cP_{C}$ enforces that recoupling condition.

If the bounded region is a single vertex $v$ of the spin network, and thus the surface is simply the set of edges attached to that vertex, then the closure defect vanishes, $C_{v}=0$. This is the intertwiner condition that ensures the gauge invariance of the spin network states. However, as shown in \cite{Livine:2013gna}, as soon as the bounded region is made of several vertices, the closure defect $C$ can be non-trivial and accounts for the possible excitations of the curvature within the region (defined as non-trivial holonomies around loops inside the region).
And it was shown in \cite{Freidel_Livine_2010,Livine_2013} that each Hilbert space $\cR^{(N)}_{J,C}$, at fixed total area $J$ and closure defect $C$, carries an irreducible representation of the unitary group $\U(N)$.

Of course, for a generic open surface, that is not necessarily a closed surface bounding a region of space, there are absolutely no constraints on the closure defect .

\medskip

We could proceed directly from that kinematical description of quantum surface states and define models of dynamics on that Hilbert space. Nevertheless, in this paper, we prefer to introduce the classical phase space underlying that construction and first investigate models of classical dynamics of such discrete surface before coming back to the quantum theory.

\subsection{Spinors for Discrete Surfaces}


Considering a single surface patch, the Hilbert space $\cH=\bigoplus_{j} V_{j}$ can be seen as the quantization of a pair of harmonic oscillators, following Schwinger representation of the $\su(2)$ algebra. This has led to the spinor representation of loop quantum gravity \cite{Freidel:2010bw,Borja:2010rc,Livine:2011gp,Dupuis:2012vp}. Here we review how this provides a classical phase space description for quantum surfaces as defined above.

Considering a discrete surface made of $N$ surface patches, we introduce for each surface patch a spinor $\ket{z_i}\in\C^{2}$ with  the label $i$ running from 1 to $N$. The two components of the spinors correspond to the pair of harmonic oscillators and are provided with the canonical symplectic structure:
\be
\ket{z_i} = \begin{pmatrix} z^0_i \\ z_i^1 \end{pmatrix}
\,,\quad
\lbrace z_i^A, \overline{z}_j^B \rbrace = -\mi \delta^{AB} \delta_{ij}\,.
\label{comm_rel}
\ee
Each spinor $z_{i}$ defines a 3-vector $\vX_{i}$ by projecting it onto the Pauli matrices:
\be
\vX_{i}\equiv\la z_{i} |\vsigma| z_{i}\ra\,,
\quad
X_{i}\equiv|\vX_{i}|= \la z_{i}| z_{i}\ra\,,
\ee
$$
| z_{i}\ra \la z_{i}|=\f12\left(
X_{i}\id+\vX_{i}\cdot\vsigma
\right)\,.
$$
We also introduce the dual spinors $|z_{i}]$ obtained by acting on the original spinor with the $\SU(2)$ structure map:
\be
|z_{i}]
=
\eps\,|\bz_{i}\ra
=
\mat{c}{-\bz^1_{i}\\\bz^0_{i}}\,.
\ee
These dual spinors give the opposite 3-vectors:
\be
[ z_{i} |\vsigma| z_{i}]=-\vX_{i}
\,,
\quad
[ z_{i} | z_{i}]= \la z_{i}| z_{i}\ra\,.
\ee

The 3-vectors defined by the spinors are interpreted as the flux vectors carried by the spin network and puncturing the surface. Geometrically, they give the normal vectors to the surface. Moreover a spinor carries one extra degree of freedom compared to the 3-vector. It is the phase of the spinor and encodes the extrinsic curvature angle in the twisted geometry interpretation of spin networks \cite{Freidel:2010aq,Freidel:2010bw}.

\begin{figure}[!ht]
\label{surface}
\begin{center}
	\begin{picture}(200,120)
		\put(0,0){\includegraphics[width=0.4\textwidth]{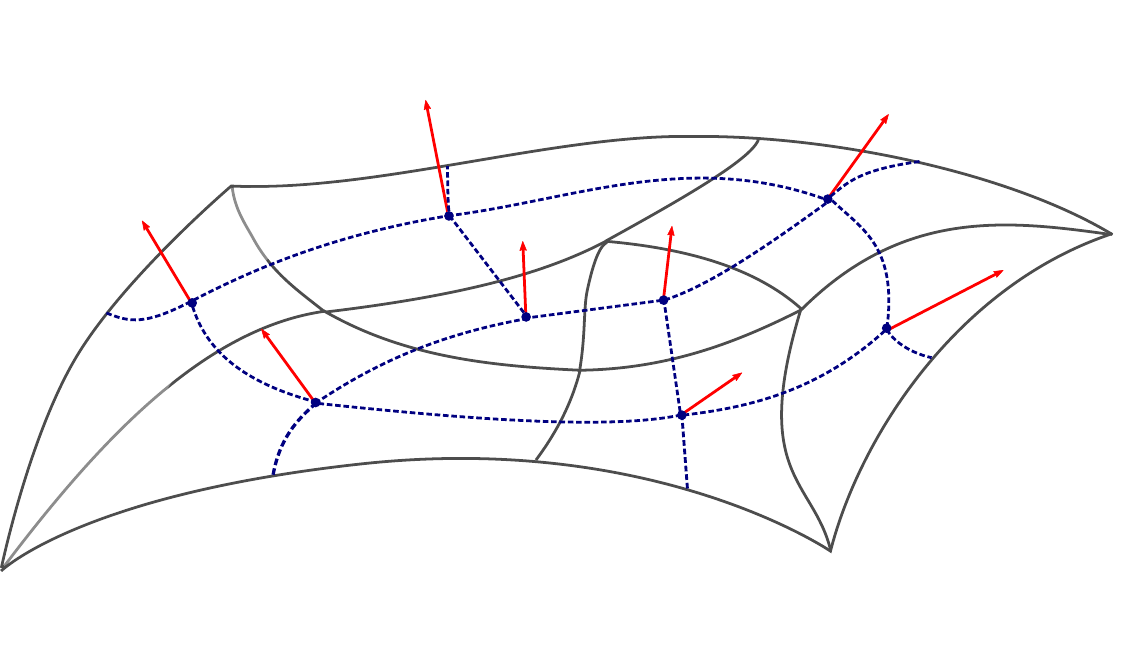}}		
		 \put(170,40){$\cS$}
		 \put(70,110){$\ket{z_i}$} \put(150,110){$\ket{z_j}$}
	\end{picture}
 \caption{\label{surface_lqg}
 A surface $\cS$ constructed from $N$ patches is defined as a collection of $N$ spinors $\ket{z_i}$. The spinor defines the normal vector to each patch, with the norm $\langle z_i | z_i \rangle$ giving the area of the patch $i$. We further define a notion of locality on the surface as a graph (in dotted blue lines) linking nearest neighbor surface patches, thought of as induced by the spin network state encoding the quantum state of the bulk 3d geometry.}
\end{center}
\end{figure}
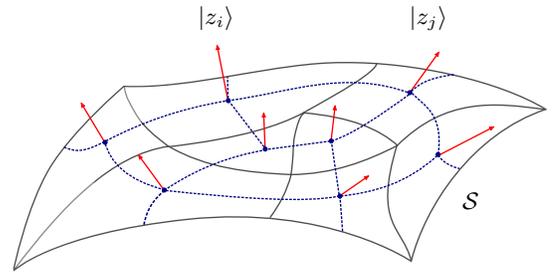

So, as drawn on fig.\ref{surface_lqg}, a discrete surface is described classically as set of $N$ (flat) surface patches, each defined by its corresponding normal vector $\vX_{i}$, whose norm $X_{i}$ gives the area of the surface patch and whose direction $\hX_{i}\in\cS_{2}$ is orthogonal to the surface (plane). The surface total area is given by the sum of all those norms:
\be
A_{\cS}=\sum_{i}X_{i}=\sum_{i}\la z_{i}|z_{i}\ra\,.
\ee

The components of each normal vector form a $\su(2)$ Lie algebra:
\be
\{X_{i}^a,X_{i}^b\}=2\eps^{abc}X_{i}^c\,.
\ee
They generate $\SU(2)$ transformation on the corresponding spinor. Now summing all those 3-vectors defines the closure defect associated to the surface:
\be
\vC
=\sum_{i}\vX_{i}
= \sum _i \langle z_i | \vsigma | z_i \rangle\,.
\ee
It generates global $\SU(2)$ transformations acting simultaneously on all the spinors:
\be
\{C^a,C^b\}=2\eps^{abc}C^c\,,
\ee
\begin{align}
\lbrace \vC , \ket{z_i} \rbrace = \mi \vsigma \ket{z_i}, \quad \me ^{\lbrace \vec{u}.\vC, . \rbrace } \ket{z_i} = g \ket{z_i}
\end{align}
with $g=\me^{\mi \vec{u}.\vsigma} \in \SU(2)$. Geometrically these are 3d rotations of the surface, rotating the normal vectors $\vX_{i}$ by a global $\SO(3)$ transformation.

Both the total area and closure defect are encoded in a single Hermitian matrix $\cX$:
\be
{\cal X}
=\sum_i \ket{z_i} \bra{z_i}
= \f12\left(A \,\id+{\vC \cdot \vsigma}\right)
\,,
\ee
where the vector $\vC$ is seen as a Bloch vector.

In the simplest case, when the bulk region is made of a single vertex (or more generally, when the spin network within the region is a tree, without loops) and the surface is simply the dual surface surrounding that vertex, the spinors satisfy the closure constraints:
\be
\vC=\sum_{i}\vX_{i}=0\,.
\ee
These constraints are first class and generate the $\SU(2)$ gauge transformations acting at the vertex. In that case, the symplectic quotient of the spinor phase space by the closure constraints $\C^{2N} //\SU(2)$ defines the phase space of framed polyhedra 
with $N$ faces, that convex polyhedra in the Euclidean 3d space up to 3d rotations and translations and with an extra phase in $\U(1)$ attached to each face (defining a local 2d frame) \cite{Freidel:2009ck,Freidel:2010aq,Freidel:2010tt,Livine_2013}. Those phases are the canonical variables conjugate to the faces' areas.

As soon as the region is composite, containing several vertices and at least one closed loop, the vector $\vC$ does not vanish anymore and we have a closure defect. Then the two basic global observables describing the discrete surface are the area $A$ and the closure vector $\vC$, which can be understood respectively as its monopole and dipole moments. When investigating global dynamics of the surface, it is natural to check how it affects these two geometric observables.

Finally, in order to go back to the quantum theory, each spinor is to be quantized as a spin and we recover spin networks and quantum surfaces defined as a collection of spins: we proceed to the canonical quantization of the spinor components, the 3-vector components $X^{a}_{i}$ become the $\su(2)$ generators $J^{a}_{i}$ acting on the elementary surface $i$ and the norm $X_{i}=|\vX_{i}|$ becomes the spin  carried by that surface patch and giving its quantized area.

\subsection{$\SU(2)$-Observables and Surface Deformations}

We can identify a generating set of $\SU(2)$-invariant observables, as defined in \cite{Girelli:2005ii,Freidel:2009ck,Freidel:2010tt,Livine:2013tsa}, given by the scalar products between the spinors and their dual:
\begin{align}
E_{ij} = \langle z_i | z_j \rangle = \overline{E_{ji}}\,,
\quad
F_{ij} = \left[ z_i | z_j \right\rangle = - F_{ji}\,.
\end{align}
These observables commute with the closure vector:
\be
\{\vC,E_{ij}\}=
\{\vC,F_{ij}\}
=0\,,
\ee
which implies that they are invariant under the global $\SU(2)$ action on the discrete surface, or equivalently under global 3d rotations\footnotemark{}.
\footnotetext{
For instance, one can recover the vector scalar products from the spinor scalar products:
$$
\big{|}E_{ij}\big{|}^{2}= 
\f12\big{(}
X_{i}X_{j}+\vX_{i}\cdot\vX_{j}
\big{)}
\,,\quad
\big{|}F_{ij}\big{|}^{2}= 
\f12\big{(}
X_{i}X_{j}-\vX_{i}\cdot\vX_{j}
\big{)}
\,.
$$
}
These natural set of observables define the basic local deformation modes of the surface: the Hamiltonian flow generated by the Poisson bracket with $E_{ij}$ or $F_{ij}$ acts on the spinors $z_{i}$ and $z_{j}$ and deforms the corresponding elementary surface patches.

More precisely, the $E_{ij}$ commute with the total area $A=\sum_{i}E_{ii}$ and define the area-preserving deformation modes. They form a closed $\u(N)$ algebra:
\be
{\{}E_{ij},E_{kl}\}= -\mi \left(\delta_{kj}E_{il}-\delta_{il}E_{kj} \right)\,.
\label{E_un}
\ee
They generate finite $\U(N)$ transformations, which allow, for instance, to explore all polyhedra with $N$ face and fixed total boundary area from any initial configuration \cite{Livine:2013tsa}.

The $F_{ij}$ and their complex conjugate $\overline{F_{ij}}$ define the basic non-area-preserving deformation modes. Since the $F_{ij}$'s are holomorphic, they are moreover invariant under $\SL(2,\C)$ transformations (as complexified $\SU(2)$ transformations). The observables $F_{ij}$ decrease the total area while $\overline{F_{ij}}$ increase it. At the quantum level, they become annihilation and creation operators, which can be used to define coherent states for intertwiners and surface states \cite{Freidel:2010tt,Livine:2013tsa,Dupuis_2010}. Together with the $\u(N)$ observables $E_{ij}$, they form a closed algebra under the Poisson bracket:
\begin{subequations}
\begin{align}
{\{}E_{ij},F_{kl}\} =& -\mi \left(\delta_{il}F_{jk}-\delta_{ik}F_{jl}\right),\\
{\{}E_{ij},\bF_{kl}\} =& -\mi \left(\delta_{jk}\bF_{il}-\delta_{jl}\bF_{ik}\right), \nn\\
{\{} F_{ij},\bF_{kl}\}=& -\mi \left(\delta_{ik}E_{lj}-\delta_{il}E_{kj} -\delta_{jk}E_{li}+\delta_{kl}E_{li}\right),  \nn\\
{\{} F_{ij},F_{kl}\} =& 0,\qquad {\{} \bF_{ij},\bF_{kl}\} =0. \nn
\end{align}
\end{subequations}
This algebra has been identified in \cite{Girelli:2017dbk} as the $\so^{*}(2N)$ Lie algebra\footnotemark{}, generating the Lie group $\SO^{*}(2N)$.
\footnotetext{
The non-compact Lie group $\SO^{*}(2N)$ has $\U(N)$ as its maximal compact subgroup and is most easily represented as a subgroup of the symplectic group $\Sp(4N,\R)$.
}
This non-compact group describes all the (linear) deformation modes of the discrete surface that commute with the closure defect (and which are thus  $\SU(2)$-invariant).


\subsection{Bulk-induced Locality as Surface Graph}
\label{SurfGraph}

We have introduced the observables $E_{ij}$ and $F_{ij}$ and described them as local deformation modes of the surface. This only makes sense if there is a notion of locality on the surface, for which the pair of surface patches $i$ and $j$ are close. In the continuum, the bulk metric for the near-by 3d geometry induced a 2d metric on the boundary surface. In the discrete setting of loop quantum gravity, we expect the spin network state, which encodes the quantum state of the 3d geometry, to induce a notion of locality on the discrete boundary surface.

Following this logic, we postulate a graph structure on the surface: we represent the elementary surface patches as nodes and draw links between nearest neighbors. Then one can use this notion of nearest neighbors to define local interactions on the discrete surface (see fig.\ref{surface_lqg} for an illustration), through for instance $E_{ij}$ or $F_{ij}$ operators acting on nearest neighbor surface patches.

\begin{figure}[!ht]
\label{surface}
\begin{center}
{\includegraphics[width=0.3\textwidth]{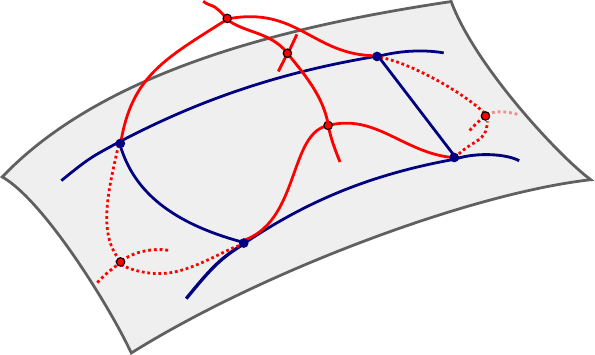}}		
 \caption{\label{surface_lqg2}
The spin network links (in red) puncture the surface and are projected onto it. They defines a surface graph with links between nearest neighbor surface patches: two patches are neighbors if the corresponding spin network links meet at a vertex in the bulk.
 }
\end{center}
\end{figure}

One possible definition of this surface graph is to project the spin network state onto the surface: one draws a link between surface patches if the corresponding spin network edges meet at a spin network vertex in the bulk (on either side of the surface), as shown on  fig.\ref{surface_lqg2}. Technically, this means that a holonomy operator acting on a loop starting at that bulk vertex and going along those two edges will act and deform that pair of surface patches. This minimalistic definition is probably naive, but it provides a first concrete proposal for the notion of bulk-induced locality on a quantum surface in loop quantum gravity.
%

As an example, in the simplest case that the bulk region contains a single spin network vertex, all the surrounding surface patches are considered as nearest neighbors: the surface graph is completely connected and all the observables $E_{ij}$ or $F_{ij}$ are legitimate local deformation modes. As soon as the bulk region will contain more spin network vertices, the notion of locality induced on the boundary surface will have to be refined and the connectivity of the surface patches will decrease.

\section{Global Surface Dynamics}
\label{GlobalDyn}

A surface in quantum gravity is not an isolated system. It is an object that lives in the 3d space and that is in constant interaction with the 3d geometry and its degrees of freedom. So we need to envision the surface dynamics as a system in interaction with the bulk geometry thought of as its environment. At the classical level, this leads to the possibility of a dissipative dynamics, not necessarily encoded as a Hamiltonian dynamics. And at the quantum level, this would lead to decoherence phenomena (see e.g. \cite{Feller:2016zuk}). 

In this section, we will focus on models of global dynamics of the quantum surface. This is meant to be especially relevant in the context of coarse-graining loop quantum gravity: the surface bounds a region of the 3d space which is coarse-grained to a single vertex, so that the surface degrees of freedom are considered as described an effective dressed vertex of a coarse-grained spin network state. So we focus on the main two global geometric observables: the area and the closure defect.
This leads us to two basic models.
First, we present a dissipation model where the closure defect will relax to the closure constraint while the area decreases to a minimal value, somehow defining a notion of ``rest area'' for the surface \cite{Livine:2013gna}.
Second, we present a forced rotation model, with a precession of the closure defect.
Both models are to be thought of as effective dynamics induced by the bulk dynamics interacting with the surface.

\subsection{Dissipation towards the Closure Constraint}

Let us place ourselves in the coarse-graining scenario for loop quantum gravity: a region bounded by a closed surface is to be coarse-grained to a single vertex. In general, the composite nature of the region leads to a non-trivial closure defect for the boundary surface. This closure defect accounts for possible curvature within the region's bulk. This is in contrast with the description of a single vertex of a spin network, which enforces a closure constraint at the vertex. That closure constraint actually ensures that we can embed the local 3d geometry around each vertex into the flat 3d Euclidean space.

Here, we introduce a model of effective dynamics\footnotemark, which relaxes a non-trivial closure defect back to the closure constraint. It can be understood as  erasing and flattening the potential curvature excitations which have built in the bulk.
\footnotetext{
We do not attempt to describe the microscopic evolution of the bulk given by some exact quantum gravity dynamics, and we focus on the effective dynamics induced on the boundary.
}
We will implement this through a continuous Lorentz transformation, boosting along the direction of the closure defect, asymptotically leading back to the position at rest with a vanishing closure defect.

\smallskip

Considering a discrete surface made of $N$ patches, we recall the Hermitian matrix $\cX$ encoding both area and closure defect:
$$
\cX=\sum_{i}|z_{i}\ra\la z_{i}|=\f12\left(
A+\vC\cdot\vsigma
\right)\,,
$$
and introduce its traceless component:
\be
\tcX=\cX-(\tr \cX)\,\frac{\id}{2}
=\f12\,\vC\cdot\vsigma\,.
\ee
With these notations, we define a first order equation of motion\footnotemark:
\beq
\label{dissip_dynamics}
\pp_{t}\,|z_{i}\ra
&=&
-\gamma\tcX\,|z_{i}\ra\\
&=&
-\gamma
\sum_{j} \la z_{j}|z_{i}\ra \, |z_{j}\ra
+\f\gamma 2\sum_{j}\la z_{j}|z_{j}\ra \, |z_{i}\ra
\,,\nn
\eeq
with an arbitrary real parameter $\gamma\in\R$ setting the relaxation speed. These are non-linear evolution equations. And as we see from the definition of $\tcX\propto \vC\cdot\vsigma$, the only fixed point of the evolution is when the closure defect vanishes, in which case the time derivatives vanish too.
\footnotetext{
To be a bit more general, we can add a proper dynamics contribution with  oscillation frequencies $\omega_i$ for each spinor:
\be
\pp_{t}\,|z_{i}\ra
\,=\,
\mi \omega_{i}\,|z_{i}\ra
\,-\gamma\tcX\,|z_{i}\ra
\,.\nn
\ee
which reduces to \eqref{dissip_dynamics} by a change of variable 
$\ket{z_i} \rightarrow \me^{\mi \omega_i t } \ket{z_i}$. This works because the matrix $\cX$ is invariant under  phase shifts of the spinors.
}

This is  a dissipative non-Hamiltonian dynamics.
%
%
Nonetheless, such an evolution could be rephrased as a high damping limit of (second order) Hamiltonian dynamics. Actually, this model is quite similar in spirit and in practice to the Kuramoto model for synchronisation \cite{pikovsky2001}, whose simple first order dynamics can be recast as an extreme dissipation limit of a more standard Hamiltonian dynamics. In this context, we could envision that our closure defect dynamics result from the interaction of the surface with a thermal bath  of bulk degrees of freedom. We postpone this to future investigation.

\smallskip

Let us integrate the equations of motion \eqref{dissip_dynamics} and obtain the explicit evolution of the discrete surface. This is not obivous, because the matrix $\tcX$ depends on the spinors themselves and the equations of motion is non-linear. We start by identifying constants of motion. It turns out that the $\SU(2)$-invariant observables $F_{ij}=[z_{i}|z_{j}\ra$ all remain constant during the evolution. To prove this, we first compute the equation of motion for the dual spinors\footnotemark{}:
\be
\pp_t |z_{i}] =\eps\,\pp_{t}|\bz_{i}\ra=-\gamma\,\eps\,\overline{\tcX}\,\eps^{-1}\,|z_{i}]
\,=\gamma\,\tcX\,|z_{i}]\,.
\ee
\footnotetext{
In order to compute the evolution of the dual spinor, we use the following identity true for an arbitrary spinor:
$$
|z\ra \la z|+|z][z|=\la z|z\ra\,\id\,,
$$
which implies that $\eps\,\overline{\cX}\,\eps^{-1}=\tr\cX\,\id -\cX$ upon summing over all the spinors.
}
Taking into account that the matrix $\cX$ is Hermitian and so is $\tcX$, this allows to show that:
\be
\pp_t F_{ij}
=\pp_{t}[z_{i}|z_{j}\ra
=\gamma [z_{i}|\tcX|z_{j}\ra-\gamma [z_{i}|\tcX|z_{j}\ra
=0\,.
\ee
These are very strong constraints on the evolution. Indeed, as showed in \cite{Dupuis:2011wy}, if two collections of spinors, $z_{i}$ and $w_{i}$, have equal scalar products $F_{ij}$, then they are equal to each other up to a global Lorentz transform $\Lambda\in\SL(2,\C)$:
\beq
&&\forall i,j\,,
[z_{i}|z_{j}\ra=[w_{i}|w_{j}\ra \\[3pt]
&&\,\Longrightarrow\,
\exists \Lambda\in\SL(2,\C)
\,\textrm{s.t.}\,
\forall i\,, |w_{i}\ra =\Lambda |z_{i}\ra\,.\nn
\eeq
This implies that the evolution of the spinors is entirely given by Lorentz transformations $\Lambda(t)\in\SL(2,\C)$ acting on their initial values:
\be
\forall i,\,|z_{i}(t)\ra\,=\Lambda(t)\,|z_{i}(t=0)\ra\,.
\ee
This Lorentz transformation does not depend on the label $i$ but acts globally on all the spinors. This leads to the evolution of the matrix $\cX$, which contains both the area and the closure defect:
\be
\cX
=\f12\left(
A\id +\vC\cdot\vsigma\right)
=\Lambda \cX_{0} \Lambda^{\dagger}\,,
\ee
where we write $\cX_{0}=\cX(t=0)$ for the initial condition. In particular, its determinant $\det \cX=\det\Lambda \cX_{0} \Lambda^{\dagger}=\det\cX_{0}$ is constant since $\SL(2,\C)$ matrices have unit determinant. This gives an essential constant of motion\footnotemark:
\be
\pp_{t}\big{[}\det \cX\big{]}
=
\f14\pp_{t}\big{[}A^{2}-\vC^{2}\big{]}=0
\,\,
\Rightarrow
A^{2}-\vC^{2}
=
A_{\infty}^{2}
\ee
where the notation $A_{\infty}$ will be justified below. Furthermore, the 4-vector $(A,\vC)$ actually transforms as a relativistic vector under the $\SO(3,1)$ transformation defined by $\Lambda$.
\footnotetext{
Seeing that $\det\cX=\sum_{i,j}|F_{ij}|^{2}=(A^{2}-\vC^{2})/4$ is a constant of motion, it is tempting to take it as a Hamiltonian, $H=\f\beta2\det\cX$ with a coupling $\beta>0$. This Hamiltonian is positive and we easily compute its flow:
\be
\pp_{t}|z_{i}\ra
=\{H,|z_{i}\ra\}
=-\mi\beta \sum_{j}F_{ij}|z_{j}\ra
=-\mi\beta(\cX-A\id)\,|z_{i}\ra\,.
\nn
\ee
There are subtle differences with the dissipative dynamics that we consider. Indeed, the coupling $\mi \beta$ is purely imaginary, hinting to an oscillatory behavior and not a dissipative process.
This means that it will not relax asymptotically to the the closure constraint.
Actually the whole matrix $\cX$ is invariant during the evolution so that both $A$ and $\vC$ are constants of motion.
Moreover, the scalar products $F_{ij}$ are not constant of motion anymore but their phases oscillate, $\pp_{t}F_{ij}=i\beta A F_{ij}$.
Finally, a  vanishing closure defect is not a fixed point of this Hamiltonian dynamics: when $\vC=0$, we are left with a phase oscillation of the spinors (with frequency given by the total area), which nevertheless leave the flux vectors $\vX_{i}$ invariant and thus almost defines a stationary regime.
}

To get the explicit evolution, we can compute the equation of motion for the full matrix $\cX$:
\begin{align}
\pp_t \cX &= \sum_i \pp_t \ket{z_i} \bra{z_i} + \ket{z_i}\pp_t \bra{z_i} \nonumber \\
	&= - \gamma \tcX\cX - \gamma \cX \tcX^\dagger = -2\gamma \cX \tcX\,.
\end{align}
Decomposing this equation onto the identity and the Pauli matrices gives the equations of motion for the area and the closure defect:
\begin{subequations}
\begin{align}
\f12\pp_{t}A &= -\gamma\vC^{2}=\gamma(A_{\infty}^{2}-A^{2})
\\
\f12\pp_{t}\vC &= -\gamma A\,\vC\,.
\end{align}
\end{subequations}
These coupled non-linear equations can be solved using the constancy of $(A^{2}-\vC^{2})$. The solution is the area converging to its asymptotic value $A_{\infty}$ as a hyperbolic tangent and the closure defect exponentially vanishing while remaining parallel to its initial value $\vC_{0}$:
\begin{subequations}
\begin{align}
A(t) &= A_\infty \text{cotanh}\left( \f t{\tau_r} + \alpha_0 \right) \\
\vC(t) &= \frac{A_\infty}{\sinh\left( \f t{\tau_r}+ \alpha_0 \right)}\,\f{\vC_{0}}{|\vC_{0}|}
\end{align}
\end{subequations}
for positive times $t\ge 0$, with $\alpha_{0}>0$ giving the initial area at $t=0$ and $\tau_r^{-1} = {2 \gamma A_\infty}$ defining the characteristic relaxation time.
This relaxation time $\tau_{r}$ becomes  shorter as the damping rate $\gamma$ is taken high or the   area large.
%
\begin{figure}[!ht]
\begin{center}
\includegraphics[width=0.8\columnwidth]{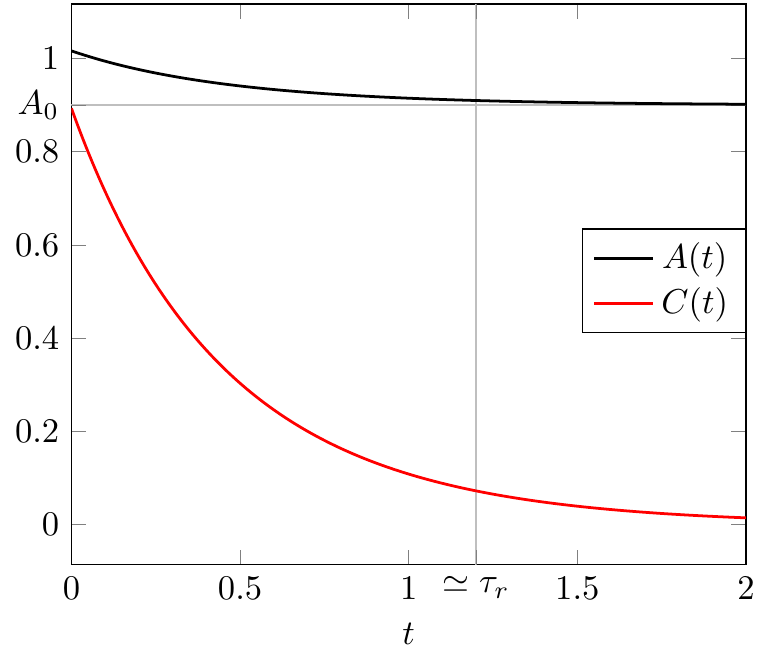}
\caption{Evolution in time of the area and closure defect: the closure defect $C\equiv|\vC|$ relaxes toward the closure constraint $C\rightarrow 0$ whereas the area relaxes toward the rest area $A_\infty$ in a characteristic time $\tau_r$.
 \label{relaxation}}
\end{center}
\end{figure}

Starting from an initial configuration $(A_{0},\vC_{0})$ with a non-trivial closure defect, the evolution acts as a Lorentz transformation on the relativistic 4-vector $(A,\vC)$ boosting it back asymptotically to its rest configuration $(A_{\infty}, \vec{0})$ with a vanishing closure defect and a rest area $A_{\infty}=\sqrt{A_{0}^{2}-\vC_{0}^{2}}<A_{0}$, as shown on the plots on fig.\ref{relaxation}. The Lorentz transformation can be made explicit, re-writing the evolution in terms of a boost rapidity $\eta$:
\be
\left\{
\begin{array}{lcl}
A&=&A_{\infty}\cosh\eta\,, \\
|\vC|&=&A_{\infty}\sinh\eta\,,
\end{array}
\right.
\quad
\eta=
\ln\left( \frac{1+\me^{-\alpha_{0}}\me^{-\frac{t}{\tau_r}}}{1-\me^{-\alpha_{0}}\me^{-\frac{t}{\tau_r}}}\right)\,
\ee
\be
\Lambda=e^{\f\eta2\,\hat{C}_{0}\cdot\vsigma}\in\SL(2,\C)
\quad\textrm{with }
\hat{C}_{0}=\f{\vC_{0}}{|\vC_{0}|}\,.
\ee
The Lorentz transformation $\Lambda$ is a pure boost along the direction of the closure defect, and its rapidity $\eta$ asymptotically vanishes in the late time limit $t\rightarrow\infty$, allowing to recover  the closure constraint $\vC\rightarrow 0$.
This relaxation to the rest frame is entirely a dissipative process.

To summarize, the spinors flow along a $\SL(2,\C)$ orbit from an arbitrary non-trivial closure defect back to a rest frame satisfying the closure constraint. The total boundary area also evolves towards its minimal rest area.
In the context of coarse-graining spin networks, this dissipative flow erases the curvature excitations within the region's bulk.

\subsection{Precessing the Closure Defect}

Let us imagine a slight variation of the previous model, coupling the spinors to their dual:
\be
\label{dynam_dual}
\pp_{t}\,|z_{i}\ra
\,=\,
-\gamma\tcX\,|z_{i}]
\,.
\ee
If the closure constraint is satisfied, $\vC=0$, then the matrix $\tcX$ vanishes too and we have a fixed point. However, this fixed point will not 
be attractive. Actually, this system does not relax to the closure constraint and we get a forced rotation motion of the closure defect. This precession explores a completely complementary regime to the dissipative model studied above.

In order to analyze the model's dynamics, we need to look deeper into the geometric interpretation of the spinors living on each surface patch. Each spinor $z_{i}\in\C^{2}$ defines the surface normal vector $\vX_{i}$, but it further defines a whole orthonormal basis in $\R^{3}$:
\begin{subequations}
\begin{align}
\vX_i =&\la z_i|\vsigma|z_i\ra \\
\vY_i =&\f12\,\big{(}\la z_i|\vsigma|z_i]+[ z_i|\vsigma|z_i\ra\big{)} \\
\vZ_i =&\f i2\,\big{(}\la z_i|\vsigma|z_i]-[ z_i|\vsigma|z_i\ra\big{)}
\end{align}
\end{subequations}
where the resulting three vectors $\vX_{i}, \vY_{i},\vZ_{i}$ have the same norm and are orthogonal to each other. 
While $\vX_{i}$ is the projection of the Hermitian matrix $|z_{i}\ra\la z_{i}|$ on the Pauli matrices, $(\vY_{i}-\mi \vZ_{i})$ is the projection of the traceless matrix  $|z_{i}]\la z_{i}|$.

We introduce the sum over all surface patches of those three vectors. The sum $\vC=\sum_i\vX_{i}$ being the closure defect, while we get two new vectors, $\vY=\sum_i\vY_i$ and $\vZ=\sum_i\vZ_i$. Actually we can repackage all these, together with the total area, in terms of 2$\times$2 matrices, by introducing the traceless matrix $\cY=\sum_i \dket{z_i}\bra{z_i}$:
\be
\cX=\f12(A\id + \vC\cdot\vsigma)
\,,\quad
\cY=\f12(\vY-\mi \vZ)\cdot\vsigma\,.
\ee
Although, for each spinor $z_{i}$, the triplet of vectors $(\vX_{i}, \vY_{i},\vZ_{i})$ is an orthonormal basis, the three vectors $(\vC,\vY,\vZ)$ are a priori not orthogonal to each other. We nevertheless have the inequalities that their  norms $C,Y,Z$ are all less or equal to the total area $A$.

\smallskip

Let us assume that the parameter $\gamma$ is real. The general case of a complex coupling, together with the whole details on the derivation and solution of the equations of motion, can be found in the appendix \ref{Twisting regime_annex}. For $\gamma\in\R$, the equations of motion read:
\be
\pp_{t}\vC=\gamma\vZ\times\vC
\,,\quad
\pp_{t}A=-\gamma\vY\cdot\vC\,,
\ee
\be
\pp_{t}\vY=-\gamma A\vC\,,\quad
\pp_{t}\vZ=0
\,.
\ee
The direction $\vZ$ remains constant during the evolution. The closure defect $\vC$ rotates around the $\vZ$ direction and no relaxation occurs.
This precession dynamics is exactly the same as a spin in a constant  magnetic field. $\vZ$ plays the role of the effective magnetic field and $\vC$ the role of a classical spin. In that context, such dynamics would be obtained from an Hamiltonian $H\propto\vC\cdot\vZ$, but the present model is more intricate and carries more degrees of freedom and can not be derived from this Hamiltonian\footnotemark.
\footnotetext{Let us consider the Hamiltonian flow generated by
$$
H\equiv \gamma\vC\cdot\vZ = \f{\mi\gamma}2\,\tr \cX(\cY-\cY^{\dagger})\,.
$$
It almost leads to the equations of motion \eqref{dynam_dual} that we postulated for the spinors:
$$
\{H,|z_{i}\ra\}=-\gamma\tcX\,|z_{i}]+\f\gamma2(\cY^{\dagger}-\cY)\,|z_{i}\ra\,.
$$
}
%
\begin{figure}[!ht]
\label{precession}
\scalebox{.8}{
\begin{picture}(150,130)
\put(0,0){\includegraphics[height=4cm]{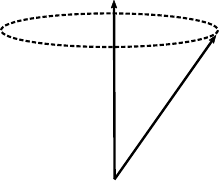}}		
\put(68,120){$\vZ$} \put(140,85){$\vC$} 
\end{picture}}
 \caption{We are in a precession regime: the closure defect defining the global polarization of the surface rotates around a constant axis, set by $\vZ$.}
\end{figure}

Let us turn to the evolution of the area. Its fate is coupled to the vector $\vY$ and we have a new invariant of motion, $(A^{2}-\vY^{2})$. This points towards a possible interpretation of this model in terms of Lorentz transformations, but  we have not identified explicitly such a representation. Instead, we have looked for a closed differential equation satisfied by the area. As shown in appendix  \ref{Twisting regime_annex}, we compute the successive derivatives of the area, and consequently of the scalar product $(\vY\cdot\vC)$, and finally obtain a fourth order differential equation satisfied by the area. Its solutions have four possible modes:
\begin{align}
A(t) =
\alpha_\pm \me^{\pm \eta t} 
+ \beta_\pm \me^{\pm \mi \om t}+ \alpha_{0}
\,,
\end{align}
in terms of the roots of the corresponding quartic polynomial:
\begin{align}
\eta=\f{\gamma}{\sqrt{2}}\,\Bigg{[}
\sqrt{(C^2-Z^2)^2 + 4 (\vC\cdot\vZ)^2} + (C^2 - Z^2) 
\Bigg{]}^{\f12}
\,,
\nn
\end{align}
\begin{align}
\om=\f{\gamma}{\sqrt{2}}\,\Bigg{[}
\sqrt{(C^2-Z^2)^2 + 4 (\vC\cdot\vZ)^2} - (C^2 - Z^2) 
\Bigg{]}^{\f12}
\,.
\nn
\end{align}
Let us keep in mind that $C^{2}$, $Z^{2}$ and $(\vC\cdot\vZ)$ are all constants of motion.
Although the motion of the closure defect $\vC$ is purely oscillatory and periodic, the evolution of the area has exponential modes and oscillatory modes. The constants of integration $(\alpha_\pm,  \beta_\pm,  \alpha_{0})$ depend entirely on the initial conditions\footnotemark {} and determine which evolution modes the area actually follow.
\footnotetext{
The initial conditions are the initial values of all the spinors $z_{i}$, but actually  focusing on the evolution of the area and closure defect, we only need to focus on $(A,\vC,\vY,\vZ)$ described by 10 real parameters. We have 5 constants of motion, the vector $\vZ$, the norm $|\vC|$ and the scalar product $\vC\cdot\vZ$, plus the 5 constants of integration parameterizing the trajectory of the area. Once the evolution of $A$ is given, the trajectory of $\vY$ is entirely determined.
}
On the one hand, it is natural for the area to have an oscillatory mode, since the motion of the closure defect is also oscillatory. On the other hand, a forced rotation motion can also pulse a constant flow of energy inducing a hyperbolic trajectory. One can indeed check that both regimes are effectively realized by choosing suitable initial conditions (see in appendix \ref{RotationTrajectory} for full details).

\smallskip

To summarize, this new dynamics we have introduced proposes a complementary regime to the dissipative dynamics we defined earlier. It induces a straightforward rotation of the closure defect, without affecting its norm. If we interpret the closure defect as a measure of the local curvature or (gravitational) energy density within the region's bulk, this would model  an object or region with a rotating energy-momentum. Moreover, the evolution of the surface area has two possible modes: an oscillatory mode forced by the rotation and an exponential mode leading to a hyperbolic trajectory for the area. Such global dynamics is very likely to be relevant to models of cosmological evolution or of astrophysical objects (with the increasing area potentially describing an exploding object or region of space).

\section{Local Dynamics on Surfaces}
\label{LocalDyn}

Up to now, we have looked into global dynamics of the surface, coupling all the surface patches together to produce a global motion for the area and the closure defect associated to the overall surface.
We have explored dissipative effects and forced rotation of the surface modeling, in an effective manner,  the presence of unmonitored bulk degrees of  freedom thought of as the environment to the surface. 

In this section, we propose to investigate local dynamics on the surface, with local fluctuations of the elementary surface patches through coupling between nearest neighbors. Moreover, since we have already introduced dynamical models for the area and closure defect, we will focus here on an isolated regime, with the surface at equilibrium with constant area and closure defect. In particular, this regime would  model the dynamics of the horizon for (quantum) black holes.

The relation of nearest neighbors between surface patches is formalized as a surface graph or network, as explained earlier in section \ref{SurfGraph}. The interactions between nearest neighbors are thought as resulting from bulk operators (e.g. holonomy operators) ending on the surface or going through the surface. 
We will not attempt to explore the details of the bulk-boundary interactions and study the projection of bulk dynamics onto the boundary surface. We instead take the point of view of \textit{effective dynamics}. Using the spinor variables to describe the state of the discrete surface, we proceed to a natural polynomial expansion of the Hamiltonian in spinor variables and we analyze the physics and dynamics induced by each possible term starting from the lowest order. The surface dynamics induced by any possible regular bulk dynamics could in principle be decomposed in such a way.

We will see that at the lowest order (the quadratic order in the spinors), the general ansatz for a Hamiltonian is a Bose-Hubbard model, with an interaction between area patches and a local potential, built from the basic area-preserving deformation operators $E_{ij}$.
Quartic order terms will lead to Ising-like Hamiltonian and so on when going to higher orders.

\subsection{Fixed Area Dynamics: the Bose-Hubbard model on the Horizon}

We would like to investigate the surface dynamics in the fixed area regime. We have in mind the application to the dynamics of isolated horizons (and thus to black hole horizons). Indeed, the energy associated to a isolated horizon is directly proportional to its area  \cite{Perez_2011,Frodden:2011eb}. Taking such a simple Hamiltonian,
\be
H\equiv \kappa A=\kappa\sum_{i}E_{ii}=\kappa\sum_{i}\la z_{i}|z_{i}\ra
\,,
\ee
leads to a almost completely stationary dynamics for the surface. The normal vectors $\vX_{i}$ do not evolve. The area of each surface patch are constant, as well as the total area. The only degrees of freedom that evolve are the phases of the spinors, which oscillate at a frequency set by the coupling factor $\kappa$:
\be
|z_{i}(t)\ra\,=\, e^{\mi \kappa t}\,|z_{i}(t=0)\ra
\,.
\nn
\ee
Here we would like to go beyond this stationary regime and introduce a framework where we can study perturbations of the surface. For instance, we would like to be able to analyze how perturbations propagate on the surface of a black hole when a system is thrown through the black hole horizon or when a Hawking radiation particle evaporates from the horizon.

The natural next-to-leading order dynamics is to introduce a hopping term between surface patches. We identify nearest neighbor patches  and define a Hamiltonian realizing local area quanta exchanges on the surface. This lead to an intrinsic local dynamics on the quantum surface.

\subsubsection{Hopping Dynamics: the $\u(N)$ Hamiltonian}
\label{Fixed Area Dynamics}

The lowest order $\SU(2)$-invariant Hamiltonian, defining area-preserving local interactions between nearest neighbor on the surface graph, is quadratic in the spinors and given by a linear combination of the $E_{ij}$ observables (defined as the scalar product between spinors):
\be
H^{\{J_{ij}\}}
\,=\,
-\sum_{\langle i,j \rangle} J_{ij}\la z_{i}|z_{j}\ra
\,,
\ee
where the $J_{ij}$ are the interaction couplings between nearest neighbor patches $\langle i,j \rangle$.
Since there is no a priori reason to distinguish links on the surface graph, we work in the homogeneous case with a global coupling $J\in\R$:
\be
H
\,=\,
-J\sum_{\langle i,j \rangle} \la z_{i}|z_{j}\ra\,,
\ee
where the coupling matrix $J_{ij}$ is taken to be proportional to the surface graph adjacency matrix.

Considering nearest neighbor interactions is standard in tight binding models in condensed matter physics. In our context, the notion of locality on the surface and the  dynamics of the surface are induced by the evolution of the bulk geometry, so a more general type of
interaction could of course be envisioned. Nevertheless, working with ``block-by-block'' dynamics in the bulk, as usually in discrete geometry models for quantum gravity (in order to keep a causal evolution), leads naturally to nearest neighbor interactions on the surface. Thus this choice is not only a matter of simplicity in defining our template surface dynamics for loop quantum gravity, but it is also enough in order to identify universality classes of surface dynamics induced by causal bulk dynamics.

\smallskip

The equations of motion are straightforwardly obtained from the phase space structure $\{z_{i}^{A}, \bz_{i}^{B}\}=-\mi \delta^{AB}$,
\begin{align}
\mi\partial_t \ket{z_i} = -J\sum_{j} \alpha_{ij} \ket{z_{j}}\,,
\end{align}
where $\alpha_{ij}$ is the adjacency matrix of the surface graph. This is easily integrated as a $\U(N)$ transformation:
\be
\ket{z_i(t)}=
\big{(}e^{\mi J \alpha}\big{)}_{ij} \ket{z_{j}(t=0)}\,,
\ee
with $e^{\mi J \alpha}\in\U(N)$  is unitary since the adjacency matrix $\alpha$ is real and symmetric.
This fits perfectly with $\U(N)$ being the group of all linear area-preserving deformations of discrete surface with $N$ faces.

In condensed  matter, usually considering a regular lattice, it is convenient to take the Fourier transform of  this $\U(N)$ evolution.
For instance, for a periodic $1d$ lattice, the adjacency matrix is:
\begin{align}
\alpha_{1d}=
\begin{pmatrix}
0&1& 0 &\dots &0&1 \\
1&0& 1 & \dots &0& 0 \\
0&1& 0 & \dots &0& 0 \\
\vdots & \vdots&\vdots &\vdots &\vdots &\vdots \\
1&0&0 & \dots &1& 0\\
\end{pmatrix}\,,
\nn
\end{align}
and the equations of motion reduce to:
\be
\pp_{t }z_{k}=\mi J\,(z_{k-1}+z_{k+1})\,.
\ee
Taking the Fourier transform,
\be
\zeta_{l}=\f1{\sqrt{N}}\sum_{k}z_{k}e^{-\mi\f{2\pi lk}N}
\,,\quad
z_{k}=\f1{\sqrt{N}}\sum_{l}\zeta_{l}e^{+\mi\f{2\pi lk}N}\,,
\ee
we simplify the equations of motion to:
\be
\pp_{t}\zeta_{l}=2\mi J \cos\f{2\pi l}N \zeta_{l}\,.
\ee
A basis of solution is given by the Bloch waves,
\be
{}^{(l)}z_k^A = Z^A\me^{\mi (q_{l}k - \omega_{l} t )}\,,
\ee
for an arbitrary fixed spinor $Z$, with wavelength and frequency:
\be
q_{l}=\f{2\pi l}N
\,,\quad
\omega_{l} = -2 J \cos \f{2\pi l}N\,,
\ee
labeled by the integer $l$ running from 0 to $N$-1. Of course we have Bloch wave modes for both  components $A=0,1$ of the spinors.

Let us check the response of the system to a local perturbation and look at the evolution of a localized excitation at a site $K$.
The initial condition at $t=0$ is the set of spinors $z_{k}(t=0)=Z\delta_{kK}$ for a given spinor $Z$. Taking the Fourier transform and computing the evolution gives:
\begin{align}
z_{k}&=\f ZN\sum_{l=0}^{N-1} e^{\f{2\mi \pi l}N (k-K)-2\mi tJ \cos\f{2\pi l}N}
\end{align}
For large $N$ number of sites (i.e. of surface patches), this is approximated by a Bessel function (Riemann integral approximation of a sum),
\begin{align}
z_{k}\underset{N\to\infty}{\sim} Z J_{2(k-K)}(2tJ)
\underset{t\to\infty}{\sim}  \f Z{\sqrt{\pi t J}}\cos \left(2tJ-\f\pi4\right),
\end{align}
with $J_n$ are  the Bessel functions of the first kind.

\begin{figure}[!ht]
\includegraphics[height=4cm]{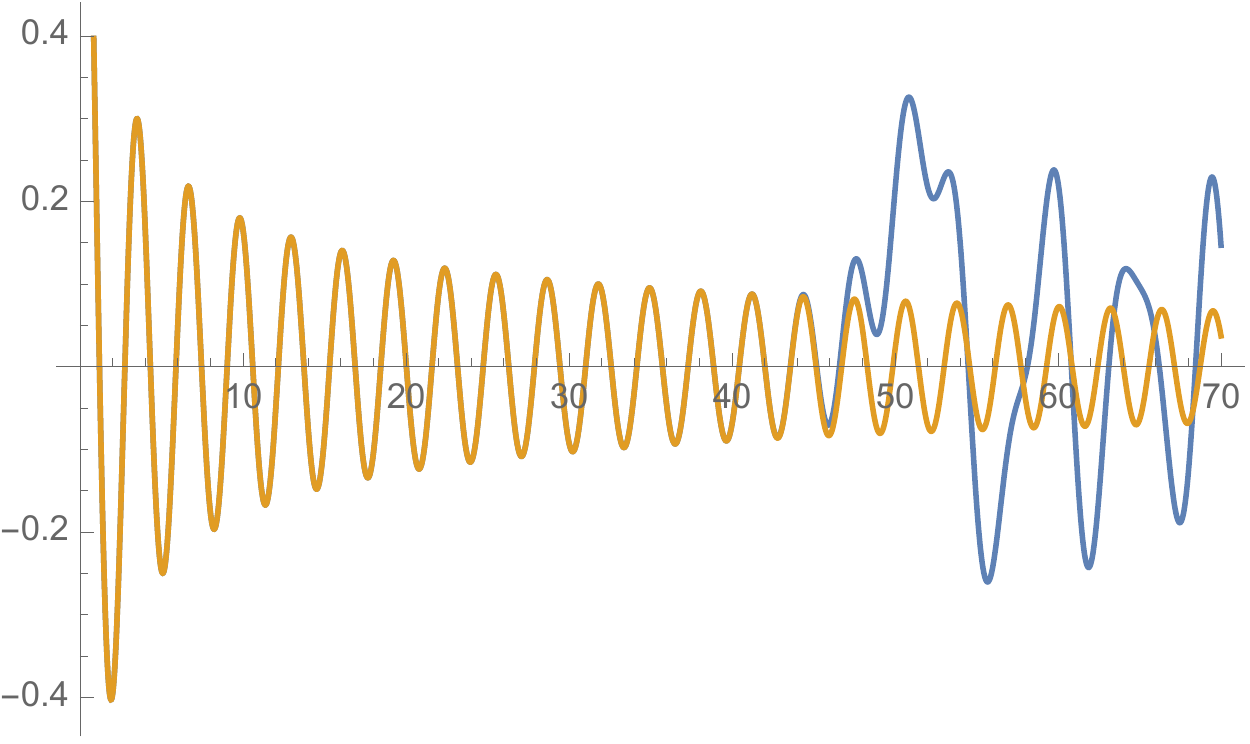}	
\caption{\label{hopplot1}
For a 1d lattice of size $N=100$, we study the propagation and diffusion of a perturbation initially localized at the site $k=0$. We look at the evolution of the amplitude at the initial site in terms of the time $tJ$. The amplitude follows very closely its Bessel approximation with oscillations decreasing in $t^{-\f12}$ until $tJ$ reaches $N/2$ and the oscillations start growing again.}
\end{figure}

\begin{figure}[!ht]
\includegraphics[height=4cm]{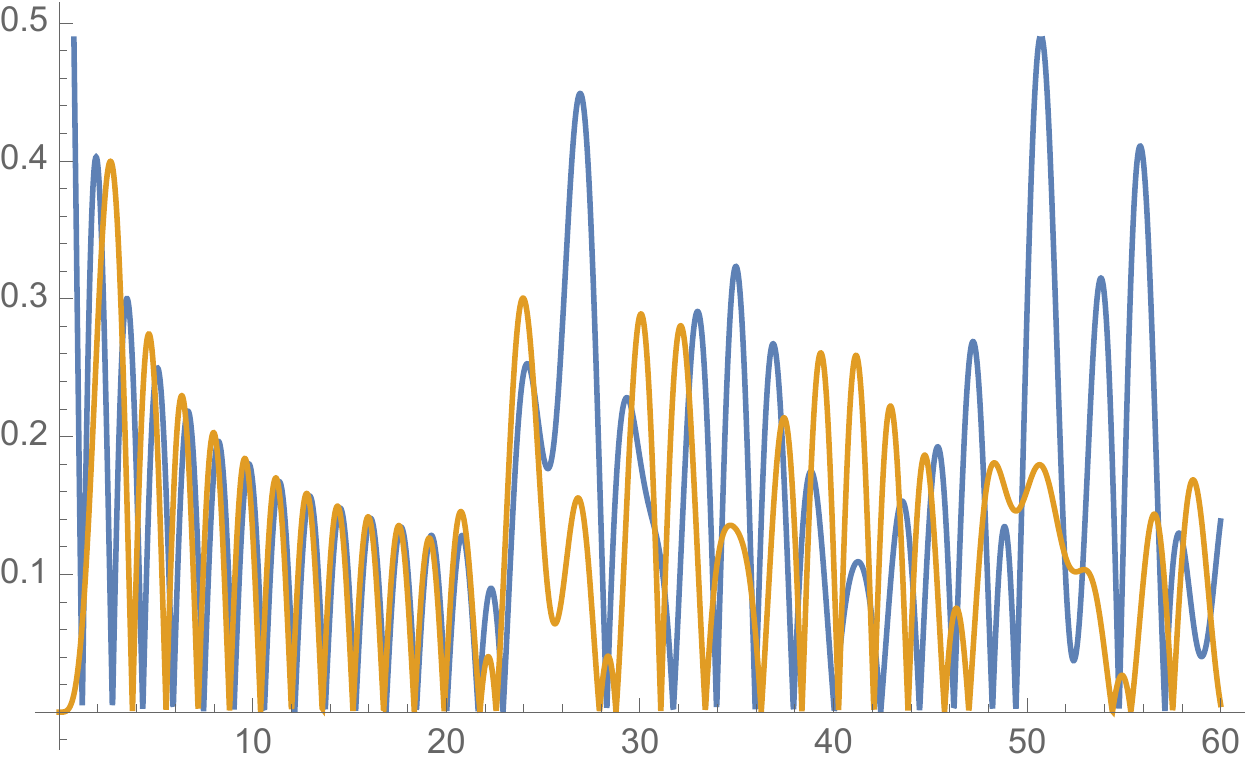}
 \caption{\label{hopplot2}
 For a 1d lattice of size $N=50$, we study the propagation and diffusion of a perturbation initially localized at the site $k=0$. We look at the amplitudes at the sites $k=0$, $k=4$ and $k=11$. At early times, the amplitudes follow their Bessel approximation. Initially the perturbation propagates until it reaches the site $k=4$ around $tJ\sim4$ and similarly for the site $k=11$. Then the amplitudes oscillate and tend to synchronize, and decrease as $t^{-\f12}$ overall. Then around $tJ\sim N/2$, this nice simple behavior breaks due to the compactness of the system.}
\end{figure}

More precisely, as long as the time $2tJ\lesssim N$ is shorter than the compact size of the 1d lattice, then the Bessel approximation describes very well the evolution of the system, as shown on fig.\ref{hopplot1}. The initial peak at $k=K$ spreads out on all the Fourier modes. At the initial site, the amplitude is maximal at $t=0$, then oscillates while decreasing in $1/\sqrt{t}$. At another site $k$, the amplitude is evanescent until the perturbation reaches it around $tJ\sim |k-K|$, at which time it reaches its maximal value before oscillating and decreasing again in $1/\sqrt{t}$, as shown on fig.\ref{hopplot2}.

\begin{figure}[!ht]
\includegraphics[height=4cm]{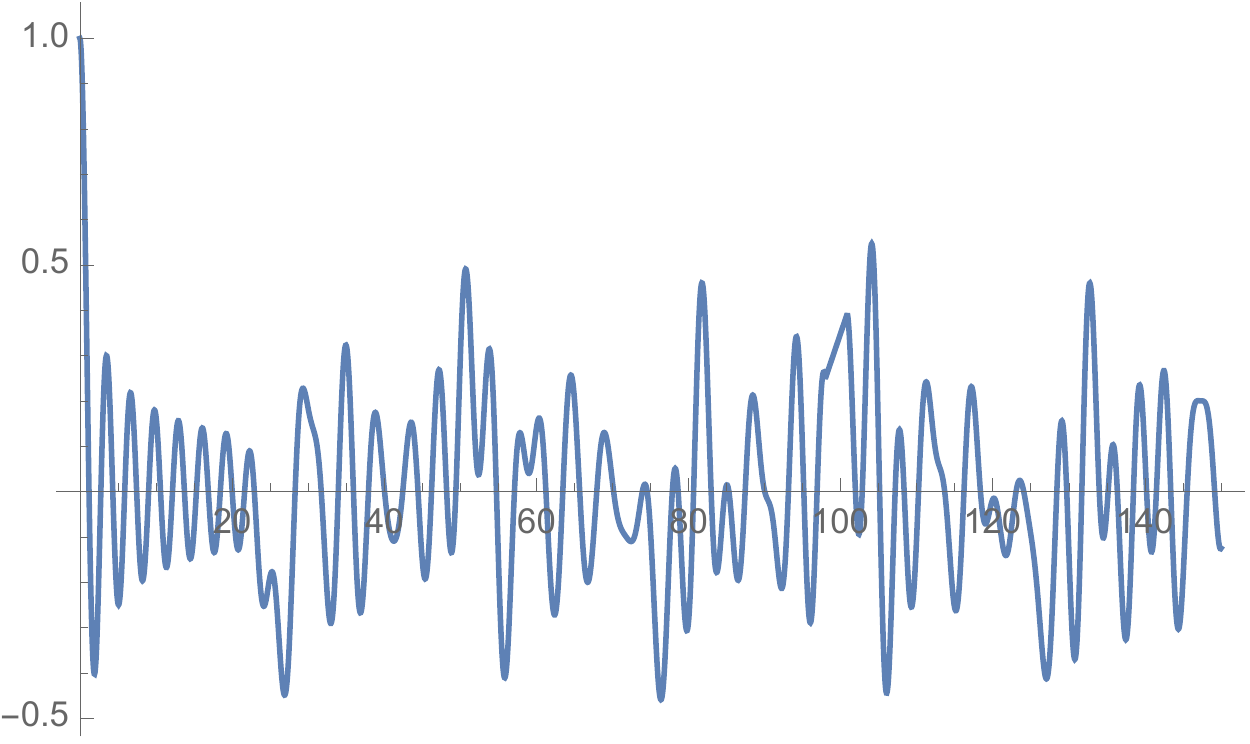}
 \caption{\label{hopplot3}
 Evolution of the amplitude at the initial site of a local perturbation at the site $k=0$ for a 1d lattice of size $N=50$ in terms of the time $tJ$.}
\end{figure}

As soon as the time is of the same order as the system size, $2tJ\sim N$, we see the effect of working on a compact lattice and the actual amplitude departs from its Bessel approximation. At the initial site, at that critical time, the amplitude's oscillations increases again and periodically reaches an almost maximal value (see fig.\ref{hopplot2} and fig.\ref{hopplot3}). We note the same behavior on all sites.

So the local perturbation excites all the Fourier modes. At early times, we do not see the finite size effects and we have a diffusive behavior. The perturbation propagates from its initial site $K$: once it reaches a given site $k$ at a time $tJ\sim |k-K|$, the amplitude keeps oscillating while  decreasing in $1/\sqrt{t}$. This way, the perturbation spreads out on the whole lattice. Then when $tJ\sim N/2$, the finite size effects kick in and we observe some kind of interference between the oscillations, with the amplitude reaching its maximal value periodically.

Here, we use a one-dimensional lattice, but our analysis would extend without complication to a regular two-dimensional  lattice.

The diffusion of the perturbation is an important feature when thinking about the application of such models of local surface dynamics to black hole horizons: it would allow the horizon to relax to equilibrium after a local perturbation, as we will discuss more in \ref{BHsection}. This comes by because the proper mode of evolution are collective wave modes. The initial local perturbation excites  all those collective modes, thus leading to its diffusion throughout the lattice until  finite size effects periodically cause some resonance and almost re-localize the perturbation. We will refine this analysis below when extending our very simple model to make it more realistic by including a local potential.

\subsubsection{Introducing a Local Potential: Bose-Hubbard Dynamics}

Up to now, we have discussed all possible quadratic terms for an area-preserving Hamiltonian: the global area and a hopping term creating area exchange between nearest neighbor on the surface. The hopping Hamiltonian can be considered as a free propagation for area degrees of freedom. Following the logic of a polynomial expansion in the spinor variables, we can supplement it with a  local potential term, quartic in the spinors (i.e. quadratic in the $E_{ij}$ and $F_{ij}$ observables). Thus we introduce a Bose-Hubbard Hamiltonian, with a local repulsion term for area quanta at each site:
\be
\label{Bose-Hubbard}
H_{BH}
\,=\,
-J \sum_{\langle i,j \rangle} \la z_{i}|z_{j}\ra\,
+\frac{U}{2}\sum_{i} \la z_{i}|z_{i}\ra^2\,,
\ee
or explicitly expanding the spinors into their two components $A=0,1$:
\begin{align}
H_{BH} = -J \sum_{\langle i,j \rangle} (\overline{z}_i^0 z_j^0 + \overline{z}_i^1 z_j^1 )
		+\frac{U}{2} \sum_i \left( |z_i^0|^2 + |z_i^1|^2 \right)^2
\,.
\nn
\end{align}
This is our main proposal of a template for the loop quantum gravity dynamics of a discrete surface of $N$ patches in the fixed area regime.
Here ``BH'' stands for Bose-Hubbard, but it will also be a proposal for the dynamics of quantum black hole horizon in loop quantum gravity.

This is a generalization of the standard  Bose-Hubbard model used in atomic physics to two coupled atomic species, since here the spinors we use here to model the surface have two independent components.

\smallskip

The equations of motion resulting from  \eqref{Bose-Hubbard} are the Gross-Pitaevskii equation on a 
lattice for each component of the spinors, 
\begin{align}
 \mi \pp_t \ket{z_i} = - J \sum_{\la k,i\ra} \ket{z_{k}} + U \langle z_i | z_i \rangle \ket{z_i}
\end{align}
This equation is usually used to describe the ground state of Bose-Einstein condensate and superfluid dynamics.

Let us come back to the case of a 1d lattice in order to illustrate simply the main features of the model. The stationary waves -the Bloch waves- obtained in the previous section when the potential vanishes $U=0$ are still stationary solutions of the Bose-Hubbard dynamics, 
\be
{}^{(l)}z_k^A = Z^A\me^{\mi (q_{l}k - \omega_{l} t )}\,,
\quad
q_{l}=\f{2\pi l}N
\ee
except that the dispersion relation is modified by the 
interaction potential,
\be
\omega_{l}
=\om_{l}^{U=0}+U\,\la Z|Z\ra
\,,\quad
\om_{l}^{U=0}
=
-2 J \cos q_{l}
\,,
\ee
with an explicit non-linear dependence on the amplitude of the wave.

\smallskip

The superfluid nature of those waves is highlighted when looking at small wave perturbations.
We look at small perturbations of the Bloch waves for the Gross-Pitaevskii equation: we choose a base mode $l$,  add a perturbation  with a slightly shifted momentum at $l\pm \delta l$ and study the stability of the mode $k$ with respect to such variation.
The calculations are detailed in  appendix \ref{BHwavestability}.

We start by looking at perturbations around the $l=0$ mode. This case, with $q_{l=0}=0$, corresponds to a homogeneous  wave $e^{-\mi \om_{0} t}$ constant in space and thus defines a homogeneous potential, equal for all surface patches. In this homogeneous case, small perturbations leads to phonon-like excitations with a Bogoliubov spectrum $\lambda_{q}\sim q v_{s}$ and a speed of sound $v_s = \sqrt{2JU\,\la Z|Z\ra}$.
This shows the stability of the waves for velocities smaller than the sound velocity. Moreover, according to Landau's criteria, the Bogoliubov spectrum ensures superfluidity: as long as an object travels in the fluid at a speed smaller than the speed of sound $v<v_{s}$, the motion is favored energetically over the excitations of perturbations and will happen without any dissipation.

On the other hand, for higher base mode $l>0$, contrary to the homogeneous case, the condensate is unstable against perturbations.  This hints new physics is involved and signals the onset of a phase transition.

\smallskip

We postpone a detailed analysis of this quantum model for surface dynamics for future investigation. It would very likely have interesting predictions for the behavior of quantum horizons in quantum gravity, their dynamics and  their phase diagram. Without performing a full analysis, we can nevertheless have a glance of what to expect by looking at the properties of the one-component Bose-Hubbard model on a regular lattice. It is a model that has been greatly studied\footnotemark{} in condensed matter and whose features  are  well-understood in atomic physics (for reviews and textbooks \cite{Krutitsky_2016,Dalibar_2008, Sachdev_book}).
\footnotetext{
Here we work in a canonical framework with a fixed area, i.e. a fixed total number of area quanta. The number of quanta at each site (on each surface patch) is an integer. To match this with the computations done in a grand-canonical framework, the chemical potential is fixed so that the average occupation number at each site $\langle n \rangle$ is an integer.
}

In particular, it  exhibits a quantum phase transition at zero temperature $T=0$ between a superfluid phase 
and a Mott phase controlled by the parameter $U/J$. We note $(U/J)_{c}$ the critical value of this ratio.
For the Bose-Hubbard model in $d$ dimensions, this transition belongs to the universality class of  the $XY$ model in $d$+1 dimensions.
As temperature is turned on, the fluid changes into a simple Bose gas phase, which can be referred to as the normal phase of the system.
\begin{figure}[ht]
%
\scalebox{.9}{\begin{tikzpicture}[domain=-2.2:2.2]

\def\xmax{8}
\def\ymax{5}
\def\xmin{0}
\def\ymin{0}
\def\xfun{\xmax*0.9}
\def\tc{0.5*\xfun}
\def\a{0.6*\ymax}

\draw[->] (\xmin,0) -- (\xmax,0) node[below] {$\frac{U}{J}$};
\draw[->] (0,\ymin) -- (0,\ymax) node[right] {$T$};

\draw[thick]  (0,\a) arc[y radius=\a,x radius=\tc,start angle=90,end angle=0] ; 
\draw[dotted]  (\tc,0) arc[y radius=2*\a,x radius=\tc,start angle=0,end angle=30]; 
\draw[dotted]  (\tc,0) arc[y radius=2*\a,x radius=\tc,start angle=180,end angle=150]; 
\draw[dashed]  (\tc,0) arc[y radius=\a,x radius=3*\tc,start angle=180,end angle=130]; 								

\node[below] () at (\tc,0) {$\left(\f{U}{J}\right)_c$};
\node[below] () at (\tc/2,\a/2) {$\text{Superfluid}$};
\node[below] () at (1.75*\tc,\a/2) {$\text{Mott}$};
\node[below] () at (\tc,1.5*\a) {$\text{Bose gas}$};

\node[circle,fill,inner sep=1.5pt]() at (\tc, 0) {} ;

\end{tikzpicture}
}
\caption{\label{BHphasediag}Phase diagram of the Bose-Hubbard model for a fixed integer average number of quanta per site.  }
\end{figure}
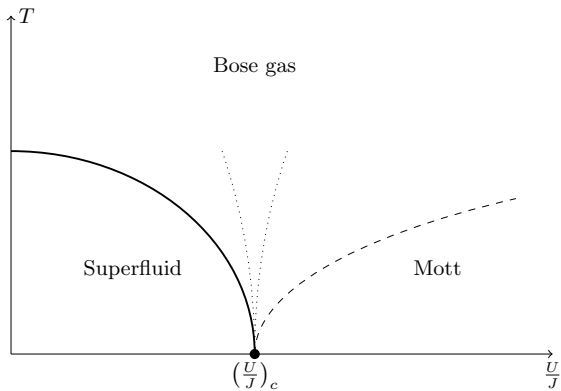

Working on a square lattice, the critical line in $d$ dimensions is given by\footnotemark
\be
\f{k_B T_c^{(d)}}{J}
=
A^{(d)}\,\left[
\left(\f UJ\right)_c - \f UJ
\right]^{z\nu}
\,,\quad
A^{(d=2)}\sim\f12\,,
\ee
where $z$ is the  node connectivity of the lattice ($z=2d$ for a square lattice in $d$ dimension) and $\nu$ is the critical exponent of the correlation length of the  $XY$ model, $\xi\sim\delta^{-\nu}$. The critical exponent is $\nu=\f12$ near the critical point for all dimensions $d>2$, except in two dimensions $d=2$ where we have $\nu\sim\f23$.
The phase diagram is drawn below on fig.\ref{BHphasediag}.
\footnotetext{
In two dimensions $d=2$, the critical line is slightly different. When we take the limit of a vanishing potential,  $U/J\rightarrow 0$, the critical line goes back to a vanishing temperature $T\rightarrow 0$. In that weakly interacting regime, we have a Berezinsky-Kosterlitz-Thouless phase transition with:
$$
\f{k_B T_c^{(d=2)}}{J}
\underset{U\rightarrow 0}\sim
\f{4\pi \la n\ra}{-\ln2\xi\f UJ}
\,.
$$
}

In fact, only the superfluid/Bose gas phase transition is associated to a symmetry  breaking and  truly define different phases. The Bose gas and the Mott phase cannot be distinguished as such.
The Mott phase is characterized by the presence of a gap $\Delta$ and the exponential damping of thermodynamical quantities.

Coming back to surface dynamics, we see that we should expect a phase transition in the dynamics at fixed area as the temperature of the surface grows, from a superfluid phase to a Bose gas.
For instance, the response of the surface to local perturbations will be different in those phases. For the Mott phase
or the Bose gas, the propagation of perturbations is ballistic whereas it is diffusive in the superfluid phase as we have already seen in the previous section in the simple model of hopping dynamics. Indeed in the Bose gas, we can have localized perturbations that propagate through the gas as particles, while the superfluid basic excitations are collective modes on the surface.
This should have interesting applications to the physics of quantum black holes, as we discuss in the next section below.   
   
\medskip
   
We have given the properties of the Bose-Hubbard on a regular lattice, but there is no a priori restriction on the type of surface graph, defining the locality on the surface. As we have already argued, the surface graph is supposed to be induced by the structure and dynamics of the spin network state underlying the bulk geometry. If we focus on the surface and forget all knowledge of the bulk geometry, we can consider an alternative point of view: the exact phases and transitions of the Bose-Hubbard model for the surface dynamics crucially depend on the details of the surface graph. Then we infer the type of surface graph  we need from the properties of the surface we expect.

For instance, for a symmetric and smooth surface, as we expect for a black hole horizon, it seems natural to consider that the surface graph is a regular lattice. But in fact, nothing \emph{a priori} forbids to consider random graphs instead
of regular ones.
From the condensed matter perspective, this amounts to introduce disorder in the system. Disorder typically blocks the diffusion of waves and leads to a localization -Anderson localization- around the defects. We speculate that this would naturally lead to the possibility of localized excitations on the surface.

More precisely, the phase diagram of the Bose-Hubbard  model is modified by disorder. A new phase called Bose glass phase appears \cite{PhysRevB.40.546}.  It is an insulating (due to localization) and a gapless phase. It would be extremely interesting for loop quantum gravity to understand the physics of the corresponding quantum surface  and what type of geometry it would correspond to.

\subsubsection{Application to Quantum Black Hole Horizons}
\label{BHsection}
 
In loop quantum gravity, the black hole horizon, as a space(-time) boundary, is pictured as a quantum surface, punctured by the spin network states defining the bulk geometry, with each puncture representing a basic surface patch and carrying quanta of area \cite{Ashtekar:2000eq}. These surface patches represent the microscopic degrees of freedom of the black hole.
This led to the paradigm of the horizon as a {\it gas of punctures}, which allows to  recover the area-entropy law \cite{Noui_2015}.

Here we propose to refine this basic picture, by introducing an intrinsic dynamics to the punctures on the horizon. Since isolated horizons are naturally in the isolated regime at fixed area, we propose to model the surface dynamics with the Bose-Hubbard model we introduced in the previous section: area quanta can now propagate along the horizon, hopping from puncture to puncture, with a repulsive local potential.

This would lead to two main predictions:
\begin{itemize}
\item A modification of the energy spectrum of the black hole, which would not be simply proportional to the area but will have corrections due to the Bose-Hubbard Hamiltonian, and which would imply corrections to the Hawking evaporation spectrum.

\item A non-trivial phase diagram for quantum black holes, with (at least) a superfluid phase and a Bose gas phase, which would depend on the notion of locality on the horizon and the surface graph induced by the near-horizon geometry.
 
\end{itemize}
We postpone a detailed study of these two predictions in the framework of the quantum Bose-Hubbard model to future work \cite{Feller_prep}, but we would already like to  outline the phase transition scenario here.

\smallskip

First of all, the choice of surface graph crucially affects the properties, and in particular the phase diagram, of the Bose-Hubbard model living on it. Thus we should identify  the phase(s) relevant to black hole physics and used this to constrain the surface graphs corresponding to a black hole horizon.

Nevertheless, considering the spherical symmetry and homogeneity of a black hole horizon, it seems natural to model it with a regular surface graph. So, assuming that we work on a (almost) regular graph, and assuming that the Bose-Hubbard couplings $J$ and $U$ are constant  (determined by the quantum gravity exact dynamics and not depending on the black hole mass or horizon area), the Bose-Hubbard phase diagram gives us is a critical temperature $T_{c}$, which depends on the ratio $U/J$ and on the (average) node valency of the surface graph. This critical temperature is to be compared to the Hawking temperature $T_{H}$ of the black hole. When the mass is large, and thus the Hawking temperature is small, we will be in the superfluid horizon phase. While when the mass is smaller and smaller, and the Hawing temperature exceeds the critical temperature, we will have a phase transition and enter the Bose gas phase.

The main difference\footnotemark{} between these two phases is how they respond to local perturbations.
\footnotetext{
Another difference between the superfluid phase and the Bose gas, which could especially be relevant to black hole physics, is the scaling law of the entropy. For instance, the entropy of a sub-region of the surface scales with its area in the Bose gas phase while it would scale with its perimeter on the ground state of the Bose-Hubbard model.
}
The superfluid phase has a diffusive behavior, perfectly suited to black holes physics, with local perturbations exciting collective modes spread out on the whole surface. On the other hand, the Bose gas phase has a ballistic behavior with local perturbations able to travel along the surface, barely deformed. So a superfluid horizon would tend to relax\footnotemark{} faster back to a homogeneous horizon after a perturbation such as an incoming mass or particle evaporation, while a Bose gas horizon would retain for a longer time the local perturbation.
\footnotetext{
Actually there can not be a true full relaxation without dissipation. If we look at a local perturbation of the horizon, due either to an incoming mass or the evaporation of a Hawking photon from the horizon, we should consider not only the intrinsic evolution of the degrees of freedom on the surface but also its coupling to the near-horizon geometry. For instance the emission of quasi-normal modes play an essential role in the relaxation of the black hole horizon to its equilibrium state with a homogeneous horizon (see e.g. \cite{Hod:2006jw}).
}

We should study the details of this scenario, identify the correct physical meaning for the Bose-Hubbard couplings $J$ and $U$,  check if we obtain reasonable values for the critical temperature, and finally if and how this phase transition scenario fits with the black hole entropy and the apparent ``information loss paradox'' in quantum gravity.

\subsection{Quartic Interactions: Ising Dynamics}

Following the logic of  a polynomial expansion of the Hamiltonian for the surface dynamics, we have already considered quadratic terms, with a local area term (which simply makes the spinors vibrate at a fixed frequency without affecting the surface geometry) and an area quanta exchange term between nearest neighbor (on the surface graph), and then a quartic local potential term, given by the square of the area of each surface patch. This has led to a Bose-Hubbard model for the surface dynamics.

We can go further and look at quartic interaction terms between nearest neighbors. Such terms turn out to give a Ising-like Hamiltonian.
Indeed we are looking for quartic real terms in the spinors living at two surface patches, say $i$ and $j$, which are $\SU(2)$-invariant and preserve the total area  (i.e. Poisson-commute with the total area). This leaves exactly only two such terms, the product of the two local areas and the scalar product of the two normal vectors:
\begin{align}
X_{i}X_{j}&=
\la z_{i}|z_{i}\ra\la z_{j}|z_{j}\ra
\\
\vX_{i}\cdot \vX_{j} &= \langle z_i | \vsigma | z_i \rangle \langle z_j | \vsigma | z_j \rangle
\\
&= 2 \langle z_i | z_j \rangle\langle z_j | z_i \rangle - \langle z_i | z_i \rangle \langle z_j| z_j \rangle \nn
\end{align}
%
%
In terms of the $E$'s and $F$'s observables, these are combination of the two quartic polynomials $E_{ij}E_{ji}$ and $F_{ij}\bar{F}_{ij}$.

If we take $i=j$, these two possible quartic terms match and are equal to local squared area $X_{i}^{2}=E_{ii}{}^{2}$, which gives the local potential of the Bose-Hubbard model.
When we consider nearest neighbors $\la i,j\ra$, the area product $X_{i}X_{j}$ seems to be a straightforward extension of the Bose-Hubbard potential and will likely not produce new physics. On the other hand, the scalar product term $\vX_{i}\cdot \vX_{j}$ is completely new. It can not be produced in a Bose-Hubbard with a single atomic species and is present in our framework because the spinors defining the surface degrees of freedom have two components. It is actually a Ising-like Hamiltonian, more precisely a Heisenberg model, corresponding to a O$(n)$-model\footnotemark{} for n=3.
\footnotetext{We recall that the O$(n)$-model for n=1 is the Ising model and for n=2 is the XY model.}

O$(n)$-models are very-well studied and exhibit phase transition between an ordered and a disordered phase. Considering such a Hamiltonian for quantum surfaces would likely lead to models where the surface patches are synchronized or dissynchronized.
And combining it with the Bose-Hubbard model would very certainly enrich the phase diagram and the physics of surfaces in loop quantum gravity.

Overall, considering all quadratic and quartic terms in the spinors, we present a full ansatz for a Hamiltonian for the dynamics of surfaces:
\be
H^{{(4)}}=
-\sum_{\la i,j\ra}
\left[\alpha \la z_{i}|z_{j}\ra +\beta \vX_{i}\cdot\vX_{j} \right]
+ \f\gamma2\sum_{i}\la z_{i}|z_{i}\ra^{2}
\ee
with a free propagation term, a coupling between nearest neighbors and a local potential. The goal is now to establish the phase diagram of these models, depending on the values of the three coupling constants $\alpha,\beta,\gamma$ and on the structure of the surface graph, and to match these coupling constants with the (loop) quantum gravity dynamics (either microscopic or coarse-grained).

\section{Conclusion \& Outlook}
\label{Conclusion}

We have looked at quantum surfaces in loop quantum gravity. Defined as a collection of elementary surface patch, a surface is further endowed with the extra structure of a graph which defines a network of nearest neighbor patches on the surface. This surface graph defines a notion of locality on the surface, thought as induced by the spin network states encoding the near-surface bulk geometry.

From this definition of a discrete surface,  we have launched the program of defining and analyzing surface dynamics in loop quantum gravity.
Describing each elementary  surface patch by a spinor at the classical level or an irreducible representation of $\SU(2)$ (spin) at the quantum level, we have introduced generic templates for both global and local surface dynamics.

\smallskip

The global dynamics focuses on the evolution of the area and closure defect of the surface due to its interaction to the bulk geometry, thought as an environment to the system. For a closed surface bounding a space region (with the trivial topology of a 3-ball), the closure defect can be interpreted as a measure of curvature in the region's bulk and vanishes by definition when the region consists in a single spin network vertex.

We explore two different models: a dissipative regime and a forced rotation regime. In the dissipative model, the surface spinors flow along a $\SL(2,\C)$ orbit and asymptotically converges back to a vanishing closure defect. In some sense, the curvature excitations within the region dissipates through the surface and the boundary area converges towards a minimal value at rest. This is the first explicit mechanism to dynamically recover the closure constraint for a composite region in loop quantum gravity.
On the other hand, the forced rotation model describes a precession motion, for which the closure defect has a constant norm and rotates around a fixed axis (defined by the initial state of the surface). The area follows a hyperbolic trajectory, together with some oscillatory modes. This could most certainly be used to model rotating astrophysical objects in the loop quantum gravity or be used in a cosmological context.

\smallskip

The local dynamics focuses on an isolated regime for which the total surface area remains fixed and describes intrinsic fluctuations of the surface geometry. We introduced a generic ansatz for a Hamiltonian, going to quartic order in the spinors (or quadratic order in the gauge-invariant observables), defined by a generalized Bose-Hubbard model (with two atomic species). It consists in a hopping term, encoding the free propagation of area quanta along the surface, plus a (repulsive) local potential (given by the squared number of area quanta). This can be supplemented with a Ising-like term favoring (or disfavoring) the alignment of the (normal) direction of neighboring surface patches.

The physics of such a Bose-Hubbard model is rich, with phase transitions between a superfluid and a Mott phase at zero temperature and between a superfluid and a Bose gas as the temperature increases. Moreover the dynamical properties of the model crucially depend on the surface graph defining the network of nearest neighbor surface patches.
This promises interesting applications to quantum black holes, with a modified energy spectrum and a possible transition from a superfluid horizon to a Bose gas horizon as the black hole becomes smaller (i.e. as its mass  decreases), which should be relevant to study in more details.

\smallskip

The next step will be to implement both global and local dynamics in the  full quantum regime of loop quantum gravity, combining the dissipative relaxation to the closure constraint to wave propagations on the surface. Not only we should analyze the phase diagrams of these models, but we should connect explicitly them to the quantum gravity dynamics, either by realizing them as effective surface dynamics in some regimes of the canonical loop quantum gravity or of spinfoam models, or at least by relating their coupling parameters (relaxation time, Bose-Hubbard couplings,...) to parameters from the full quantum gravity theory.
Then we could also extend our surface models to a variable  number of surface patches (or punctures) by introducing a chemical potential as in statistical mechanics. This would certainly lead to more realistic models for black hole horizons in a grand-canonical framework.

Finally, we believe that it is crucial to understand how the structure of the surface graph, encoding the notion of locality on the surface, affects the evolution and (thermo)dynamical properties of the quantum surface. This seems necessary, in the context of  implementating of the holographic principle, in order to clarify the conditions for a proper definition of holographic screen in loop quantum gravity, and could lead to surprises, such as (Anderson) localization or (Bose) glass phases due to disorder and randomness in the surface graph.

\section*{Acknowledgement}

We would like to thank Tommaso Roscilde from Laboratoire de Physique, ENS de Lyon,
for numerous and helpful discussions on cold atoms physics and the Bose-Hubbard model.

\appendix

\section{Integrating the Global Precession Dynamics}
\label{Twisting regime_annex}

\subsection{Equations of Motion and General Solutions}

The forced rotation model of global dynamics is defined by the following equation of motions for the spinors and their dual:
\begin{subequations}
\begin{align}
\pp_t \ket{z_i} &= -\gamma\tX \dket{z_i} \quad
\pp_t \bra{z_i} = - \gamma^* \dbra{z_i}\tilde{X} \\
\pp_t \dket{z_i}  &=-\gamma^* \tilde{X} \ket{z_i} \quad
\pp_t \dbra{z_i} - \gamma \bra{z_i} \tilde{X} 
\end{align}
\end{subequations}
In addition to the Hermitian matrix $\cX=\sum_{k}|z_{k}\ra\la z_{k}|$, we introduce another matrix:
\be
\cY
=\sum_{k} \dket{z_k} \bra{z_k}
 = \f12{\left( \vY - \mi \vZ \right)} \cdot \vsigma
\ee
The motion couples these two matrices $\cX$ and $\cY$:
\begin{subequations}
\begin{align}
\label{Xeqn}
\pp_{t} \cX&=-\gamma \tcX \cY - \bgamma \cY^\dagger \tcX \\
\label{Yeqn}
\pp_t \cY &= -\bgamma A \,\tcX \\
\pp_t \cY^\dagger &= - \gamma A \,\tcX 
\end{align}
\end{subequations}
By projecting the $\cY$-equation \eqref{Yeqn} on the Pauli matrices, we obtain the equations of motion for the vectors $\vY$ and $\vZ$:
\begin{subequations}
\begin{align}
\pp_t \vY &= - \text{Re}(\gamma) \,A\vC \\
\pp_t \vZ &= - \text{Im}(\gamma) \,A\vC
\end{align}
\end{subequations}
Combining these two equations, we  can directly conclude that the vector 
$\vR \equiv\left( \text{Re}(\gamma) \vZ - \text{Im}(\gamma) \vY \right) $ is 
a constant of motion, $\pp_t \vR = 0$.

Now projecting the $\cX$-equation  \eqref{Xeqn} on the Pauli matrices, we will obtain the equations of motion for the closure defect and the area. To this purpose, we compute:
\begin{align}
\label{trace}
\trace{\left( \tcX\cY \sigma_a \right)} &= 
	\frac{1}{4} C_b \left( Y_c-\mi Z_c \right) \trace{\left( \sigma_b \sigma_c \sigma_a \right)} \nonumber \\
	&=\frac{\mi }{2}   \epsilon_{abc} C_b \left(Y_c - \mi Z_c \right)
\end{align}
Similarly, we get:
\begin{align}
\label{trace}
\trace{\left(\cY ^\dagger \tcX \vsigma \right)} 
=
-\frac{\mi }{2}   \vC\times \left(\vY + \mi\vZ \right)
\,,
\end{align}
\begin{align}
\trace{\tcX\cY } =\f\vC2\cdot \left( \vY - \mi \vZ \right)
\,,\quad
\trace{ \cY ^\dagger\tcX } =\f\vC2\cdot \left( \vY + \mi \vZ \right)
\,.\nn
\end{align}
This allows to compute:
\begin{align}
\pp_{t}\vC
&=\f{\mi }2\,\left(
\gamma (\vY-\mi\vZ)\times\vC
-\bgamma (\vY+\mi\vZ)\times\vC
\right)\nn\\
&=\left(
-\text{Im}(\gamma) \vY
+\text{Re}(\gamma)\vZ
\right)\times\vC
=\vR\times\vC\,.
\end{align}
Since $\vR$ is constant, we get the rotation of the closure defect $\vC$ around  the $\vR$ axis at constant speed: this is a precession motion.
In particular, the norm of the closure defect $|\vC|$ is a constant of motion and never decreases unlike in the dissipative model. The scalar product $(\vC\cdot\vR)$ is also a constant of motion.

We follow the same method for the area and project the $\cX$-equation  \eqref{Xeqn} onto the identity to obtain:
\begin{align}
\pp_t A 
&=-\f12\vC\cdot\left(
\gamma(\vY-\mi \vZ)+\bgamma(\vY+\mi \vZ)
\right)
\nn\\
&= - \left( \text{Re}(\gamma) \vY + \text{Im}(\gamma) \vZ \right) \cdot\vC
\end{align}
This allows to identify another constant of motion:
\be
\pp_{t}\,\big{[}
A^{2}-\vY^{2}-\vZ^{2}
\big{]}=0\,.
\ee
We put this in contrast with the Lorentz invariant $(A^{2}-\vC^{2})$ we identified in the dissipative regime. This means that there is likely an underlying Lorentz transformation representing the motion, but we haven't followed this path to solve the equations of motion.

Integrating the equation for the area is not as straightforward as for the closure defect. Let us restrict ourselves to the case $\gamma\in\R$ for the sake of simplicity.
Then the rotation axis is $\vR=\gamma\vZ$ and the equations of motion for the area and closure defects simplify to:
\be
\pp_{t}A=-\gamma\vY\cdot\vC
\,,\quad
\pp_{t}\vY=-\gamma A\vC
\,,\quad
\pp_{t}\vC=\gamma \vZ\times \vC\,.
\nn
\ee
The vector $\vZ$, the norm $|\vC|^{2}$, the scalar product $(\vC\cdot\vZ)$ and $(A^{2}-\vY^{2})$ are all constants of motion.
To obtain a closed equation for $A$, we compute the successive differentials of the scalar product $\vY\cdot\vC$,
\begin{align*}
\pp_t \left( \vY\cdot\vC \right) &=
-\gamma A\vC^2+\gamma\vY\cdot(\vZ\times\vC)\\
\pp_{t} \left(\vZ\times\vC\right)
&=\gamma\vZ\times(\vZ\times\vC)
=\gamma(\vZ\cdot\vC)\vZ-\gamma\vZ^2\vC
\\
\pp_{t} \big{[}\vY\cdot(\vZ\times\vC)\big{]}
&=
\gamma(\vZ\cdot\vC)(\vY\cdot\vZ)-\gamma\vZ^2(\vY\cdot\vC)
\\
\pp_{t}\left(\vY\cdot\vZ\right)
&=-\gamma\,A\,\vC\cdot\vZ
\\
\pp_{t}^2\left(\vY\cdot\vZ\right)
&=\gamma^2(\vY\cdot\vC)(\vC\cdot\vZ)
\end{align*}
Keeping in mind that $\vZ$, $|\vC|$ and $(\vC\cdot\vZ)$ are all constants of motion, we get the following differential equation:
\be
\pp_{t}^2 (\vY\cdot\vC)
=
\gamma^2
\left[
(C^{2}-Z^{2})( \vY\cdot\vC)
+(\vC\cdot\vZ)(\vY\cdot\vZ)
\right]
\nn
\ee
In order to get rid of the scalar product $(\vY\cdot\vZ)$, we can apply $\pp_{t}^2$ again and get a finally closed fourth order differential equation:
\be
\Big{[}\pp_{t}^4 
-\gamma^2(C^2-Z^2)\pp_{t}^2
-\gamma^4(\vC\cdot\vZ)^2
\Big{]}
(\vY\cdot\vC)=0
\,.\nn
\ee
The solutions of this last differential equation with constant coefficients are easily found.
The discriminant of the quadratic equation,
$$
\lambda^{2}-\gamma^2(C^2-Z^2)\lambda
-\gamma^4(\vC\cdot\vZ)^2
=0
$$
is positive, $\Delta\ge 0$, and we have two roots with opposite sign, 
\begin{align}
\lambda_\pm =
\f{\gamma^{2}}2\,\Bigg{[}
{ (C^2 - Z^2) \pm\sqrt{(C^2-Z^2)^2 + 4 (\vC\cdot\vZ)^2}}
\Bigg{]}
\,.
\nn
\end{align}
One must take another square-root to take the dynamical modes of the area. The positive root $\lambda_{+}>0$ leads to exponential behavior, while the negative root $\lambda_{-}<0$ leads to oscillatory behavior:
\begin{align}
A(t) =
\alpha_\pm^{+} \me^{\pm t\sqrt{\lambda_+} } 
+ \alpha_\pm^{-} \me^{\pm \mi t\sqrt{-\lambda_- }}+ \alpha_{0}
\,.
\label{ATraj}
\end{align}
The constants of integration are functions of the initial conditions.
For instance, assuming that the coupling is positive, $\gamma>0$, if initially the vectors $\vY$ and $\vC$ are pointing in opposite directions, $\vY\cdot\vC<0$, then the area will grow and the vector $\vY$ will keep growing in the opposite direction to $\vC$, thus leading to an exponential growth. On the other hand, if the  vectors $\vY$ and $\vC$ are pointing in the same direction, $\vY\cdot\vC>0$, then the area will start by decreasing, as well as the vector $\vY$, and this would lead to either an exponential flow of the area or to an oscillatory regime, likely depending on the rotation speed and norm of $\vC$.

\subsection{Trajectory Analysis}
\label{RotationTrajectory}

Let us look deeper into the possible trajectory. The ansatz \eqref{ATraj} is general, with the exponential and oscillatory modes and the 5 constants of integration, and nothing ensures a priori that all modes are actually realized when varying the initial conditions. Moreover, one should pay special care that we have extra constraints, coming from the geometrical interpretation of $A$ as the area and the definitions of the vectors from the spinors:
\be
A\ge 0\,,
\quad
C\le A\,,
\quad
Y\le A\,,
\quad
Z\le A\,.
\ee
Our general procedure to check the realizability of a trajectory $A(t)$ is as follows. We first choose the fixed rotation vector $\vZ$, say along the $z$-azis $\vZ=\he_{z}$, where we fixed the norm of the vector to one without affecting the generality of our analysis. The trajectory of the closure defect is straightforward:
\be
\vC=v \he_{z} + u\big{(}
\cos\gamma t \he_{x}+\sin\gamma y \he_{y}
\big{)}\,,\,
C=\sqrt{u^{2}+v^{2}}
\ee
Then, if we postulate a trajectory for the area $A(t)$ plugging  values for the constants in the anstaz \eqref{ATraj}, we can entirely determine the vector $\vY$ by integrating the first order differential equation $\pp_{t}\vY=-\gamma A\vC$. Once we have the trajectory for $\vY$, we check the consistency of our trajectories by requiring that the other differential equation $\pp_{t}A=-\gamma \vY\cdot\vC$ be satisfied.
Let us also keep in ming that $(A^{2}-Y^{2})\ge 0$ is constant during the evolution, which provides yet another consistency check.

\smallskip

Let us start by considering the case for which $\vC$ is also along the direction of $\vZ$, taking the values $u=0$ and $v=C$. In that case, the roots are $\lambda_{+}=\gamma^{2}C^{2}$ and $\lambda_{-}=-\gamma^{2}$.
We first realize a hyperbolic trajectory,
\be
A=\alpha \cosh\eta
\,,\quad
\vY=-\alpha\sinh\eta\,\he_{z}+\vY_{0}\,
\ee
where the boost rapidity grows linearly, $\eta=\gamma C t +\eta_{0}$, and the constant offset $\vY_{0}$ is orthogonal to $\vC$, i.e. $\vY_{0}\cdot \he_{z}=0$. The area obviously stays positive, while the norm invariant $(A^{2}-Y^{2})=(\alpha^{2}-\vY_{0}^{2})\ge 0$, requires $Y_{0}$  to be less than the amplitude $\alpha$.

It is also straightforward to check that the oscillatory mode, $A=\alpha_{0}+\beta\cos\gamma t$, is impossible to realize in this configuration.

\smallskip

In order to realize the oscillatory regime, we explore the completely opposite configuration, taking $\vC$ orthogonal to the rotation axis $\vZ$ and choosing the values $v=0$ and $u=C$. In this case, the roots are:
\be
\lambda_{\pm}=\f{\gamma^{2}}2(C^{2}-1\pm|C^{2}-1|)\,.
\ee
We distinguish three cases. If $C>1$, then $\lambda_{+}=\gamma^{2}(C^{2}-1)$ and $\lambda_{-}=0$, so we expect a purely hyperbolic regime. If $C=1$, then the roots are both trivial $\lambda_{\pm}=0$ and the area should be simply constant. Finally, if $C<1$, $\lambda_{+}$ vanishes and $\lambda_{-}=-\gamma^{2}(1-C^{2})$, so that we expect to be in a purely oscillatory regime.

Let us place ourselves in the latter case, $C<1$, and choose an oscillatory ansatz for the area:
\be
A=\alpha_{0}+\beta\cos\om t\,.
\ee
It is straightforward to  integrate for the vector $\vY$ by simple trigonometric manipulations. Imposing both $\pp_{t}\vY=-\gamma A\vC$ and $\pp_{t}A=-\gamma \vY\cdot\vC$, we get that the oscillation frequency must satisfy:
\be
\om^{2}=\gamma^{2}(1-C^{2})=-\lambda_{-}\,,
\ee
as expected. We can also check that $(A^{2}-Y^{2})$ is still an invariant and compute its value in terms of the parameters $\alpha_{0}$ and $\beta$ of the area trajectory:
\be
A^{2}-Y^{2}=(1-C^{2})\left(\alpha_{0}^{2}-\f{\beta^{2}}{C^{2}}\right)\,.
\ee
To keep this invariant positive, we simply have to require that the constant mode $\alpha_{0}$ be larger than the oscillation amplitude $\beta/C$, which also implies that the area $A$ stays positive during the evolution.

\smallskip

In general, we expect that generic trajectories, with closure defect $\vC$ having both non-vanishing longitudinal and transversal components, will mix the oscillatory and exponential modes, although this remains to be checked explicitly.

\section{Bose-Hubbard model: Stability of Bloch waves}
\label{BHwavestability}


In the Bose-Hubbard model, used to describe the dynamics of the fluctuations of the surface geometry in the fixed area regime, let us look at the stability of the Bloch wave solutions.
We consider a  perturbation of a Bloch wave with momentum $k$ with two 
waves with momenta $k \pm q$, 
\begin{align}
\psi_j(t) =
\big{[}
\alpha
+u(t) \me^{\mi jaq} + \bar{v}(t)\me^{-\mi jaq}
\big{]}
\,
\me^{\mi\left( jak - \omega_{k}t\right)}
\end{align}
with the dispersion relation $w_{k}=-2\cos ak + U\,|\alpha|^2$.
%
We insert this ansatz in the Gross-Pitaevski equation
$\mi\hbar\pp_{t}{\psi}_j(t) = -J(\psi_{j+1} + \psi_{j-1}) + U|\psi_j|^2\psi_j$.
Expanding to first order in $u$ and $v$, we obtain a set of two linear first order differential 
equations
\begin{align}
\mi \hbar \frac{\md}{\md t} 
\begin{pmatrix}
u \\ v
\end{pmatrix}
 = \hat{O}
 \begin{pmatrix}
u \\ v
\end{pmatrix}
\end{align}
with $\hat{O}=\hat{D}+\hat{V}$  a 2$\times$2 matrix (not necessary  Hermitian),
\begin{align}
\hat{D}
&=
-2J \begin{pmatrix}
\cos a(k+q)-\cos ak & 0\\ 0  & \cos ak-\cos a(k-q)
\end{pmatrix}
\nn\\
\hat{V}&=U
 \begin{pmatrix}
|\alpha|^2 & \bar{\alpha}^2 \\ -\alpha^2 & -|\alpha|^2
\end{pmatrix}
\nn
\end{align}
The  stability is inferred by looking at the eigenvalues of this effective first order Hamiltonian $\hat{O}$, which are 
\begin{align}
\lambda^{\pm}_{q} &= 2J\sin ak\sin aq  \\
&\pm 2 \left( 4J^2\sin^4\f{aq}2\cos^2 ak + 2JU|\alpha|^2\sin^2\f{aq}2\cos ak\right)^{\f12} \nonumber
\end{align}
We distinguish several cases.
For $U=0$, we recover the Bloch dispersion relation of the free model without potential,
with
$\lambda^{\pm}_{q}=\pm{[}\om_{k+q}-\om_{k}{]}$.

For $U\ne0$, we first consider the  perturbations around the zero momentum state $k=0$. This corresponds to the homogeneous case, since $\psi_{k=0}$ does not any spatial variation and the corresponding potential is constant. Plugging $k=0$ in the eigenvalue formula above,
we recover a Bogoliubov spectrum for phonon-like excitations with speed of sound $v_s=a\sqrt{2JU|\alpha|^2}$,
%
\begin{align}
\lambda^{\pm}_{q}&=\pm2
\left[
2J\sin^2\f{aq}2\,
\left(
2J\sin^2\f{aq}2+U|\alpha|^2
\right)
\right]^{\f12}
\\
&\underset{q\rightarrow0}\sim
qv_{s}\,.
\nn
\end{align}
For small  modes $q$, the spectrum is almost linear, $|\lambda_q|\sim qv_{s}$. And we can push the analysis further. Following Landau's criteria, since we always have $q^{-1}|\lambda_q|> v_{s}$, this signals superfluidity: if an object moves slower than the speed of sound $v_{s}$ in the condensate (i.e. on the surface in our context), or equivalently if the fluid moves at a speed $v<v_{s}$, it is not favorable to create excitations (since the energy due to motion $qv$ is not enough to excite the perturbation mode $q$)  and the object/fluid can move freely without dissipation.

In the general case for $k\ne 0$, the eigenvalues can acquire a non-zero imaginary part, as soon as the ratio $U/J$ or the wave amplitude $|\alpha|^{2}$ are large enough.  This signals an instability and hints towards the existence of a phase transition.

\vfill

\bibliographystyle{bib-style}
\bibliography{LQGSurfDyn}

\end{document}